\documentclass[12pt,preprint]{aastex}

\def\pf{p_f}
\def\pr{p_r}
\def\hf{h_f}
\def\hr{h_r}
\def\rf{r_f}
\def\rr{r_r}
\def\ff{f_f}
\def\fr{f_r}
\def\bb{\beta}

\def\br{\beta_r}

\def\rhof{\rho_f}
\def\rhor{\rho_r}
\def\g43{\gamma_{43}}

\def\Gb{\Gamma}
\def\Gej{\Gamma_{\rm ej}}
\def\Eej{E_{\rm ej}}

\def\Lej{L_{\rm ej}}
\def\vej{v_{\rm ej}}
\def\bej{\beta_{\rm ej}}
\def\rhoej{\rho_{\rm ej}}
\def\rhoejRS{\rho_{\rm ej}{\rm (RS)}}
\def\rhoFS{\rho_1 {\rm (FS)}}
\def\nejRS{n_{\rm ej}{\rm (RS)}}
\def\GejRS{\Gamma_{\rm ej}{\rm (RS)}}

\def\M0{M_0}
\def\mom{\tilde p}
\def\pa{\partial}

\def\tobs{t_{\rm obs}}

\def\be{\begin{equation}}
\def\ee{\end{equation}}
\def\beq{\begin{eqnarray}}
\def\eeq{\end{eqnarray}}

\begin{document}

\title{A Semi-analytic Formulation for Relativistic Blast Waves \\
       with a Long-lived Reverse Shock}

\author{Z. Lucas Uhm\altaffilmark{1, 2}}
\affil{Institut d'Astrophysique de Paris, UMR 7095 Universit\'e Pierre et Marie Curie-Paris 6 \\
-- CNRS, 98 bis boulevard Arago, 75014 Paris, France}

\altaffiltext{1}{E-mail: uhm@iap.fr}
\altaffiltext{2}{International Center for Astrophysics, 
                 Korea Astronomy and Space Science Institute, Daejeon 305-348, Korea}

\begin{abstract}

This paper performs a semi-analytic study of relativistic blast waves 
in the context of gamma-ray bursts (GRBs). 
Although commonly used in a wide range of analytical and numerical studies,
the equation of state (EOS) with a constant adiabatic index is a poor 
approximation for relativistic hydrodynamics.
Adopting a more realistic EOS with a variable adiabatic index,
we present a simple form of jump conditions for relativistic hydrodynamical shocks.
Then we describe in detail our technique of modeling 
a very general class of GRB blast waves with a long-lived reverse shock. 
Our technique admits an arbitrary radial stratification of the ejecta and ambient medium.
We use two different methods to find dynamics of the blast wave: 
(1) customary pressure balance across the blast wave and 
(2) the ``mechanical model''. 
Using a simple example model, we demonstrate that 
the two methods yield significantly different dynamical evolutions 
of the blast wave. We show that the pressure balance 
does not satisfy the energy conservation for an adiabatic blast wave 
while the mechanical model does. 
We also compare two sets of afterglow light curves 
obtained with the two different methods.

\end{abstract}


%
%

\section{Introduction} \label{section:introduction}

The afterglow emission of a gamma-ray burst (GRB) is believed 
to be produced by a relativistic blast wave (M\'esz\'aros \& Rees 1997).
The relativistic blast wave is driven by an ``ejecta'', 
which is ejected by the central engine of the GRB explosion. 
As the ejecta interacts with a surrounding ambient medium, 
two (forward and reverse) shock waves develop (e.g., Piran 2004).
The forward shock (FS) wave sweeps up the ambient medium,
and the reverse shock (RS) wave propagates through the ejecta.

As the blast wave has high Lorentz factors $10^2 - 10^3$ (e.g., M\'esz\'aros 2006),
the FS wave is highly relativistic and an equation of state (EOS) 
with a constant adiabatic index 4/3 may well describe the gas in the FS-shocked region. 
However, the strength of the RS wave varies as the blast wave propagates. 
In the case of a constant-density ambient medium,
the RS wave is initially non-relativistic 
and then transitions to a mildly relativistic or relativistic regime
(Kobayashi 2000; Sari \& Piran 1995).
Thus, an EOS with a constant adiabatic index is not adequate 
for the gas in the RS-shocked region;
a variable adiabatic index needs to be considered 
to account for change in the gas temperature.

Although the EOS with a constant adiabatic index has been widely used 
in analytical and numerical studies of relativistic hydrodynamics,
it is valid only for the gas of either non-relativistic (with the index 5/3)
or ultra-relativistic temperature (with the index 4/3).
The correct EOS for a relativistic ideal gas is formulated 
in terms of modified Bessel functions (e.g., Synge 1957),
and its equivalent adiabatic index varies from 5/3 to 4/3 
as the temperature increases.

As it is not convenient to deal with modified Bessel functions, 
there has been effort to find simpler EOSs that closely reproduce 
the correct EOS of a relativistic ideal gas. 
Taub (1948) showed that the choice of EOS is not arbitrary 
and must satisfy a certain inequality (Taub's inequality).
By taking the equal sign in Taub's inequality, 
Mignone et al. (2005) derived a simple form of EOS that
has correct limiting values 5/3 and 4/3.

The same EOS as in Mignone et al. (2005) was previously introduced by Mathews (1971),
considering a relativistic ``monoenergetic'' gas where all particles have the same energy.
The validity of this EOS was addressed by Blumenthal \& Mathews (1976) for the cases 
of both infinite mean free collision times and very short mean free collision times. 
This EOS was also adopted by Meliani et al. (2004) and Mignone \& McKinney (2007).
In particular, Mignone \& McKinney (2007) demonstrated in their relativistic numerical 
simulations that use of an EOS with a constant adiabatic index can significantly 
endanger the solution when transitions from cold to hot gas (or vice versa) are present.

We use the same above EOS in this paper.
Following Mathews (1971), we consider a relativistic monoenergetic gas 
and show that it closely reproduces the correct EOS of a relativistic ideal gas.
Then we use this EOS to find a simple form of jump conditions for relativistic 
hydrodynamical shocks. This simple set of jump conditions applies to shocks of 
arbitrary strength.

A short-lived RS was proposed to explain a brief optical flash 
(M\'esz\'aros \& Rees 1999; Sari \& Piran 1999a, 1999b).
A dynamical evolution of such a short-lived RS was studied analytically 
by assuming an equality of pressure across the blast wave 
(Kobayashi 2000; Sari \& Piran 1995).
The RS wave here is short-lived since the ejecta is assumed to have 
a constant Lorentz factor. However, in general, the ejecta is expected 
to emerge with a range of the Lorentz factors.
The shells with lower Lorentz factors will gradually ``catch up'' with 
the blast wave as it decelerates. Thus, the RS wave is long-lived. 
Such a long-lived RS was studied for a power-law ejecta 
by assuming a constant ratio of the two pressures at the FS and RS 
(Rees \& M\'esz\'aros 1998).

In this paper, we present a detailed description of our blast-wave modeling technique 
for even more general class of explosions where the ejecta and the ambient medium 
have an arbitrary radial structure or stratification. 
More specifically, 
we study analytically the spherical expansion of such a stratified ejecta and 
find the trajectory of the RS wave through the ejecta self-consistently. 
In order to find a dynamical evolution of the blast wave, 
we use two different methods: (1) customary pressure balance and 
(2) the ``mechanical model'' (Beloborodov \& Uhm 2006).

Using a simple example model, we demonstrate that, 
although the customary assumption of pressure balance for the blast wave 
yields an estimated evolution, it is not rigorously accurate. 
In particular, the energy conservation is not satisfied 
for an adiabatic blast wave; the total energy is decreased 
by a factor of 5 in the case of the example model.

The mechanical model was developed for relativistic blast waves, 
by relaxing the pressure balance (or proportionality) and 
applying the conservation laws of energy-momentum tensor and mass flux 
on the blast between the FS and RS. 
Using the same example model, we show that
the energy conservation is satisfied for the mechanical model. 
We also show that dynamical evolutions found by the two methods 
differ significantly.
Finally, we present the afterglow light curves in X-ray and optical bands. 
We compare two sets of light curves corresponding to the two different 
dynamical evolutions mentioned above.

In Section~\ref{section:shock},
we derive a simple set of jump conditions for relativistic hydrodynamical shocks. 
In Section~\ref{section:blast_waves}, 
we describe in detail our blast-wave modeling technique.
We also provide a simple method of evaluating the blast energy, 
employing a Lagrangian description for the blast wave. 
In Section~\ref{section:mechanical_model}, 
we review the mechanical model including more detailed equations.


%
%

\section{Relativistic shocks} \label{section:shock}

We consider a shock wave of an arbitrary strength. 
The preshock medium (cold) is denoted by region 1, 
and the postshock medium (hot) by region 2. 
The gas moves at right angles to the surface of 
discontinuity (shock front).  
The rest-mass density $\rho$, the energy density $e$ (including rest energy), 
and the pressure $p$ of the gas are defined in the rest frame of each region. 
The energy-momentum tensor $T^{\alpha\beta}$ for a perfect fluid and the mass flux $j^\alpha$ 
are given as 
\be 
T^{\alpha\beta}=(e+p)u^\alpha u^\beta+g^{\alpha\beta}p 
\qquad \mbox{and} \qquad 
j^\alpha=\rho u^\alpha,
\ee
where $g^{\alpha\beta}$ is the Minkowski metric, and 
$u^\alpha$ is the 4-velocity of the gas.


\subsection{Jump conditions}

A shock is described by three jump conditions that express 
the continuity of mass, energy, and momentum flux densities, 
respectively, in the shock frame (Landau \& Lifshitz 1959),
\beq
\label{eq:mass_flux}
\gamma_2 \beta_2\, \rho_2 &=& \gamma_1 \beta_1\, \rho_1, \\
\label{eq:energy_flux}
\gamma_2^2 \beta_2\, (e_2+p_2) &=& \gamma_1^2 \beta_1\, \rho_1 c^2, \\
\label{eq:momentum_flux}
\gamma_2^2 \beta_2^2\, (e_2+p_2) + p_2 &=& \gamma_1^2 \beta_1^2\, \rho_1 c^2.
\eeq
Here subscripts 1 and 2 refer to regions 1 and 2, 
and $c$ is the speed of light.  
The Lorentz factor $\gamma$ of the gas for each region 
is measured in the rest frame of shock front, 
and thus $\beta = (1-1/\gamma^2)^{1/2}$ is the gas velocity 
relative to the shock front.  
Note that we have set $p_1=0$ and $e_1=\rho_1 c^2$ here 
since we assume that region 1 is cold.

We introduce a quantity $\kappa$ in writing an EOS 
for the relativistic gas in region 2,
\be
\label{eq:eos}
p_2 = \kappa_2\, (e_2 - \rho_2 c^2). 
\ee
We use the quantity $\kappa$ instead of an adiabatic index,
in order to avoid any confusion with the Lorentz factors.
Note that $\kappa_2$ is 2/3 for a non-relativistic shocked gas 
and 1/3 for an ultra-relativistic gas. 
As mentioned in Section~\ref{section:introduction},
we do not use a constant value for $\kappa_2$. 
The quantity $\kappa_2$ here varies between 1/3 and 2/3 
as the strength of the shock varies.

We solve these four Equations (\ref{eq:mass_flux})-(\ref{eq:eos}) algebraically. 
First, we define a compression ratio $a \equiv \rho_2/\rho_1$ 
and get the relation $\gamma_1^2 = a^2(\gamma_2^2 -1)+1$ 
by squaring Equation (\ref{eq:mass_flux}). 
Then, substituting $\gamma_1^2$ and $e_2$ (from Equation (\ref{eq:eos})) 
into Equations (\ref{eq:energy_flux}) and (\ref{eq:momentum_flux}),  
we find the expressions for $\gamma_1, \gamma_2$, and $p_2$ 
in terms of $\kappa_2$ and the compression ratio $a$,
\beq
\label{eq:gam}
&& \gamma_2^2 = \frac{(a+1)}{a(1-\kappa_2^2)+(1+\kappa_2)^2}, 
\qquad
\gamma_1^2 = \frac{(a+1)[a \kappa_2-(1+\kappa_2)]^2}{a(1-\kappa_2^2)+(1+\kappa_2)^2}, \\
\label{eq:pre}
&& p_2/(\rho_1 c^2) = \frac{(a \kappa_2)^2-(a \kappa_2)(2 + \kappa_2)}{(1 + \kappa_2)},
\eeq
where we have used the fact 
that $a(1-\kappa_2)+(1+\kappa_2)$ cannot be zero 
for $a>0$ and $1/3\leq \kappa_2 \leq 2/3$. 
Equations (\ref{eq:gam}) and (\ref{eq:pre}) are exact.

The shock strength may be described by the relative velocity 
$\beta_{12}=(\beta_1-\beta_2)/(1-\beta_1\beta_2)$, 
or by the relative Lorentz factor $\gamma_{12}=(1-\beta_{12}^2)^{-1/2}$, 
which is given by  
\be
\gamma_{12}=(1-\beta_1 \beta_2) \gamma_1 \gamma_2 
=\gamma_1 \gamma_2 - [(\gamma_1^2-1)(\gamma_2^2-1)]^{1/2}.
\ee 
The shock strength $\gamma_{12}$ is then directly calculated 
using Equation (\ref{eq:gam}),
\be
\gamma_{12} = \frac{(a\kappa_2-1)}{(1+\kappa_2)}.
\ee
Thus, we find a very simple form for the compression ratio $a$ 
in terms of $\kappa_2$ and $\gamma_{12}$,
\be
\label{eq:comp}
a = \frac{\rho_2}{\rho_1} = \frac{(1+\kappa_2) \gamma_{12}+1}{\kappa_2}.
\ee
Substituting the compression ratio (\ref{eq:comp}) 
into Equations (\ref{eq:gam}) and (\ref{eq:pre}), 
we now express $\gamma_1, \gamma_2$, and $p_2$ 
in terms of $\kappa_2$ and $\gamma_{12}$,
\beq
\gamma_1^2 &=& \frac{(\gamma_{12}+1)[(1+\kappa_2)\gamma_{12}-\kappa_2]^2}
                  {(1-\kappa_2^2)\gamma_{12}+(1+\kappa_2^2)}, \\
\gamma_2^2 &=& \frac{(\gamma_{12}+1)}{(1-\kappa_2^2)\gamma_{12}+(1+\kappa_2^2)}, \\
\label{eq:press2}                  
p_2 &=& (\gamma_{12}-1)[(1+\kappa_2)\gamma_{12}+1]\, \rho_1 c^2.
\eeq
As we shall see below, $\kappa_2$ here is in fact a function of $\gamma_{12}$,
and therefore the shock strength $\gamma_{12}$ becomes 
the only free parameter of the shock.
Combining Equations (\ref{eq:eos}), (\ref{eq:comp}), and (\ref{eq:press2}), 
we verify an expected relation,
\be
\label{eq:expec}
e_2 = \frac{1}{\kappa_2}\, p_2 +\rho_2 c^2= \gamma_{12}\, \rho_2 c^2,
\ee
which means that the shock strength $\gamma_{12}$ is equal to the mean random 
Lorentz factor of particles in the postshock medium (measured in its fluid frame).

It may be noticed that there are only three independent equations 
among Equations (\ref{eq:comp})-(\ref{eq:expec}), 
since we had three jump conditions in the beginning. 
In fact, three of them are equivalent to those equations 
that appear on Blandford \& McKee (1976).


\subsection{Relation between $\kappa$ and mean Lorentz factor $\bar \gamma$} \label{section:kappa_Lorentz}

\subsubsection{Relativistic ideal gas}

We briefly review a relativistic ideal gas 
where a particle of mass $m$ and momentum $\mom$ has the energy 
$\epsilon(\mom) = mc^2\, \left[1+\left(\mom/mc \right)^2 \right]^{1/2}$.
The Maxwellian momentum distribution function $f(\mom)$ is given as 
$f(\mom) = A\, \mom^2 \exp\left[-\epsilon(\mom)/(k_B T) \right]$, 
where $A$ is a proportionality constant, 
$k_B$ is the Boltzmann constant, 
and $T$ is the temperature of the gas. Using a normalization 
condition on the number density, $n=\int_0^{\infty} f(\mom) d\mom$, 
the proportionality constant is found 
to be $A=\frac{n}{(mc)^3}\, \frac{u}{K_2(u)}$, 
with a modified Bessel function $K_2$. 
Here we have defined $u \equiv (mc^2)/(k_B T)$, 
which basically measures how relativistic the gas is; 
$u \gg 1$ corresponds to a non-relativistic limit 
and $u \ll 1$ to an ultra-relativistic limit. 
We evaluate the integrals of pressure and energy density of the gas:
\beq
p &=& \frac{1}{3} \int_0^{\infty} \mom\, v(\mom) f(\mom) d\mom = n k_B T, \\
\label{eq:energy_ideal}
e &=& \int_0^{\infty} \epsilon(\mom) f(\mom) d\mom 
   = (n\, mc^2) \left( \frac{K_1(u)}{K_2(u)} + \frac{3}{u} \right),
\eeq
where $v(\mom)$ is the velocity of a particle of momentum $\mom$, 
and $K_1(u)$ is also a modified Bessel function. 
See Greiner et al. (1995) for an alternative derivation.
Equation (\ref{eq:energy_ideal}) gives 
the mean Lorentz factor $\bar \gamma$ of particles as 
\be
\bar \gamma = \frac{K_1(u)}{K_2(u)} + \frac{3}{u}.
\ee
The quantity $\kappa$ (defined in Equation (\ref{eq:eos})) is given as 
\be
\kappa_i = \frac{p}{e-\rho c^2} 
         = \left[ u \left( \frac{K_1(u)}{K_2(u)} + \frac{3}{u}-1 \right) \right]^{-1}.
\ee
Here the subscript $i$ refers to a relativistic {\it ideal} Maxwellian gas. 
Since both $\bar \gamma$ and $\kappa_i$ are given in terms of $u$ only, 
one should be in principle able to express $\kappa_i$ 
as a function of $\bar \gamma$. 
However, it is not easy to do so analytically, 
as it involves two modified Bessel functions $K_1$ and $K_2$. 
This becomes a good motivation of considering 
the following monoenergetic gas.


\subsubsection{Monoenergetic gas}

\begin{figure}
\begin{center}
\includegraphics[width=10cm]{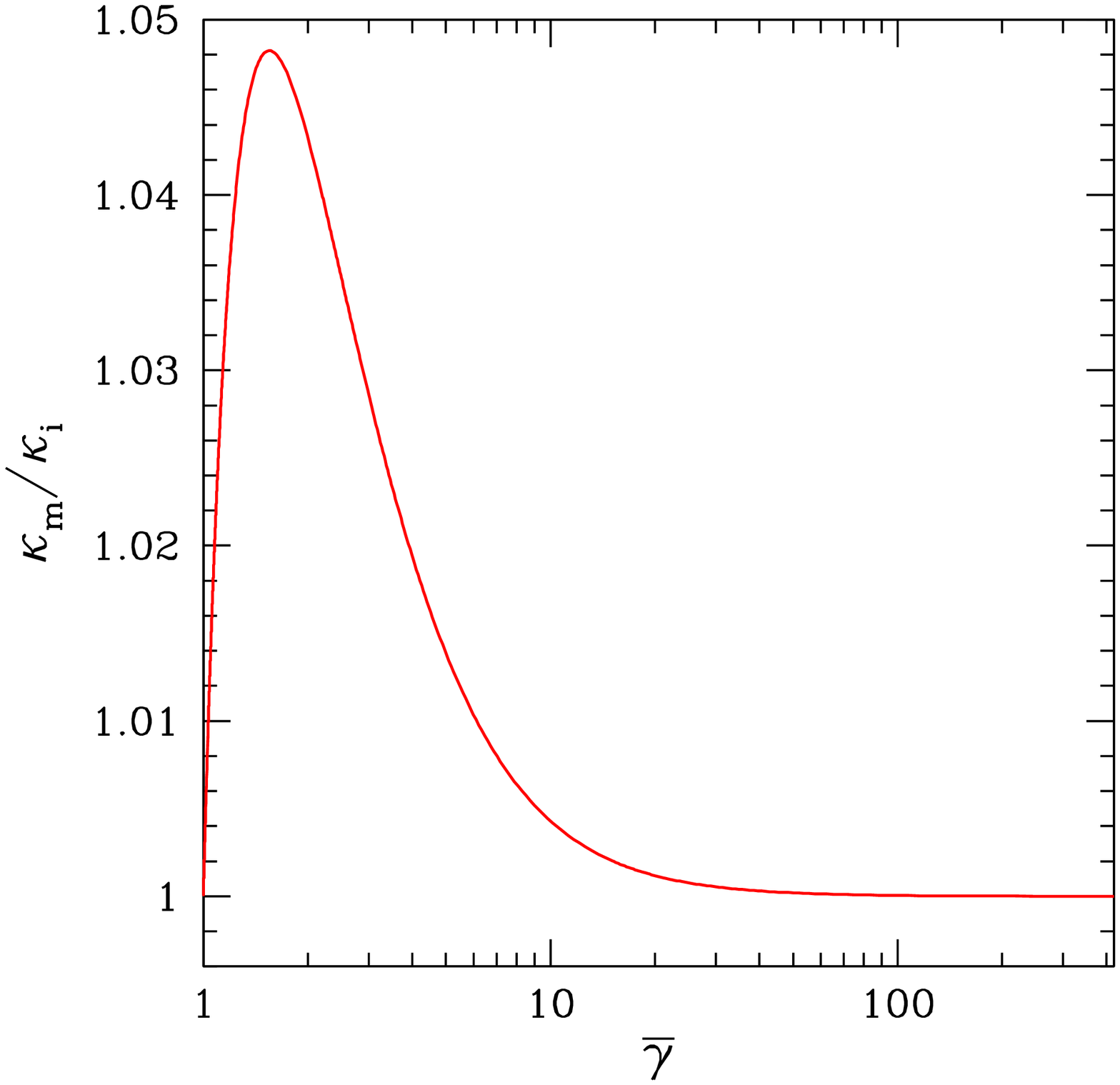}
\caption{
Ratio $\kappa_m/\kappa_i$ as a function 
of the mean Lorentz factor $\bar \gamma$ of gas particles.
The quantity $\kappa$ is defined in Equation (\ref{eq:eos}), 
an equation of state (EOS) of a relativistic gas. 
The subscripts {\it i} and {\it m} refer to 
a relativistic ideal gas and a monoenergetic gas, respectively. 
An expression for each $\kappa$ ($\kappa_i$ or $\kappa_m$) is derived 
in Section~\ref{section:kappa_Lorentz}. 
Note that there is only 4.8 \% of maximal difference between the two 
at about $\bar \gamma = 1.6$. 
} 
\label{fig:kk}
\end{center}      
\end{figure}

Here we consider a monoenergetic gas, where all particles 
in the gas have the same momentum $\bar p$ or the same Lorentz 
factor $\bar \gamma$ (e.g., Mathews 1971). 
We show below that the behavior of this gas is very close to that of a relativistic 
ideal gas. The momentum distribution function is simply given by a Dirac-delta function,
$f(\mom) = n\, \delta(\mom-\bar p)$, which satisfies the normalization condition, 
$n = \int_0^{\infty} f(\mom) d\mom$. 
We evaluate the integrals of pressure and energy density for this gas:
\be
\label{eq:pe_mono}
p = n\, (mc^2) \frac{{\bar \gamma}^2-1}{3 \bar \gamma}, 
\qquad
e = n\, (\bar \gamma mc^2).
\ee
Therefore, the corresponding $\kappa$ is found to be
\be
\label{eq:mono}
\kappa_m = \frac{p}{e-\rho c^2} = \frac{1}{3} \left( 1+\frac{1}{\bar \gamma} \right),
\ee
where the subscript $m$ refers to a {\it monoenergetic} gas. 
Note that 
$\kappa_m$ has the correct limiting value $2/3$ for a non-relativistic gas 
and $1/3$ for an ultra-relativistic gas.
We compare this monoenergetic gas with a relativistic ideal gas, 
by computing $\kappa_m / \kappa_i$ 
numerically as a function of $\bar \gamma$. 
The result is shown in Figure~\ref{fig:kk}.
Note that there is only 4.8 \% of maximal difference 
at about $\bar \gamma = 1.6$, and two gases are practically identical 
especially for high Lorentz factors above 10.

Relation (\ref{eq:mono}) is simple, 
and thus very useful in dealing with a relativistic gas.
In the following section, we show that 
the jump conditions (\ref{eq:comp})-(\ref{eq:expec}) simplify significantly 
when the gas is treated as monoenergetic.


\subsection{Jump conditions of a monoenergetic gas}

We continue on the problem of a relativistic shock wave 
for the case of a monoenergetic gas. 
When the gas in the postshock medium (region 2) 
is treated as monoenergetic, 
the quantity $\kappa_2$ satisfies 
\be
\label{eq:kap2}
\kappa_2 = \frac{1}{3} \left(1+\frac{1}{\gamma_{12}} \right),
\ee
since the mean Lorentz factor in Equation (\ref{eq:mono}) 
is equal to the shock strength $\gamma_{12}$ 
as shown in Equation (\ref{eq:expec}).
Using relation (\ref{eq:kap2}), 
we rewrite the jump conditions (\ref{eq:comp})-(\ref{eq:expec}) 
in terms of $\gamma_{12}$ only:
\beq
\label{eq:jump1}
\gamma_1^2 &=& \frac{(4 \gamma_{12}^2-1)^2}{8 \gamma_{12}^2+1}
\qquad \mbox{or} \qquad
\beta_1=\frac{4\beta_{12}}{\beta_{12}^2+3}, \\
\gamma_2^2 &=& \frac{9 \gamma_{12}^2}{8 \gamma_{12}^2+1} 
\qquad \mbox{or} \qquad
\beta_2=\frac{\beta_{12}}{3}, \\
\label{eq:jump_press}
p_2 &=& \frac{4}{3}\, (\gamma_{12}^2-1)\, \rho_1 c^2, \\ 
\label{eq:jump4}
a &=& \rho_2/\rho_1 = 4 \gamma_{12},
\qquad 
e_2 = 4 \gamma_{12}^2\, \rho_1 c^2.
\eeq
It has become clear that the shock strength $\gamma_{12}$ 
is the only free parameter of the shock.
We emphasize that these simple 
equations (\ref{eq:jump1})-(\ref{eq:jump4})
are exact for a monoenergetic gas 
and apply to shocks of arbitrary strength 
(relativistic, mildly relativistic, or non-relativistic).
This result is also briefly described in Beloborodov \& Uhm (2006).

%
%

\section{Blast waves} \label{section:blast_waves}

\begin{figure}
\begin{center}
\includegraphics[width=10cm]{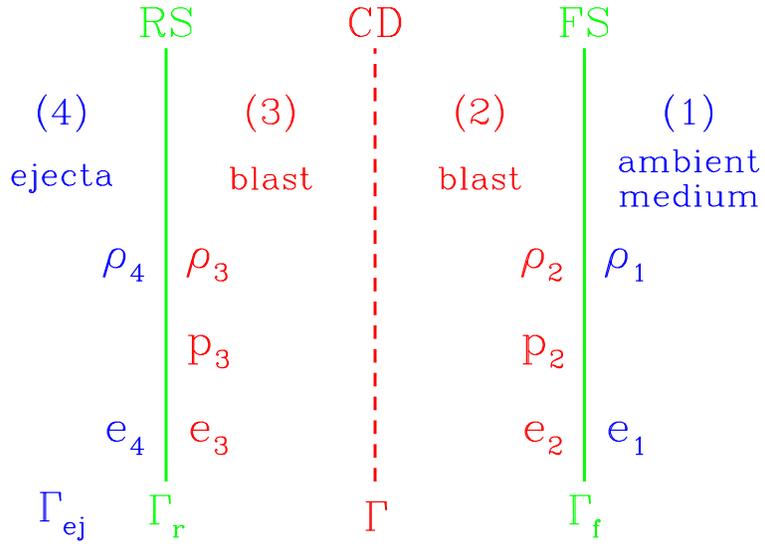}
\caption{
Illustrative diagram of 4 regions in a blast wave. 
The forward shock (FS) sweeps up the external ambient medium (region 1) 
and the reverse shock (RS) propagates through the ejecta (region 4). 
The shocked ambient medium (region 2) is separated from 
the shocked ejecta (region 3) by a contact discontinuity (CD). 
Two shocked regions 2 and 3 between the FS and RS 
are ``hot'' and called the blast.  
The pre-shock regions 1 and 4 are ``cold,'' having no pressure.
The entire blast is assumed to have a common Lorentz factor $\Gb$.
} 
\label{fig:schem}
\end{center}      
\end{figure}

A central explosion of a GRB 
ejects a large amount of material 
with high Lorentz factors $\Gej \sim 10^2-10^3$. 
This ejected flow is called the ``ejecta''.
The ejecta expands and drives a forward shock (FS) wave 
into the external ambient medium.
When the ejecta interacts with the ambient medium, 
another shock wave -- a reverse shock (RS) -- develops 
and propagates through the ejecta. 
Thus, this standard picture has four regions: 
(1) external ambient medium, 
(2) shocked external medium, 
(3) shocked ejecta, and 
(4) unshocked ejecta (e.g., Piran 2004). 
In Figure~\ref{fig:schem}, we show schematically these four regions.  
The shocked external medium is separated from 
the shocked ejecta by a contact discontinuity (CD). 
Two shocked regions 2 and 3 
between the FS and RS are hot and called the ``blast''.

We assume that the whole blast moves 
with a common Lorentz factor $\Gb$. 
This is reasonable since internal motions in the blast are subsonic, 
and hydrodynamical simulations confirm 
that $\Gb\approx \mbox{const}$ between the 
FS and RS (Kobayashi \& Sari 2000).
The Lorentz factors $\Gamma_f$, $\Gamma_r$, and $\Gej$, 
denoting for the FS, RS, and ejecta, respectively, 
are measured in the lab. frame. 
The ambient medium is at rest in the lab. frame. 
The rest-mass density $\rho$, energy density $e$ 
(including rest energy), and pressure $p$ in each region 
are measured in its own rest frame. 
It is assumed that the ambient medium and ejecta 
are ``cold,'' having no pressure.


\subsection{Radially stratified ejecta} \label{section:ejecta}

The explosion ejecta is viewed as a sequence of shells 
that coast with Lorentz factors $\Gej$. 
Each shell is prescribed an ejection time $\tau$.
The ejection time plays the role 
of Lagrangian coordinate that labels the shells in the ejecta. 
Theoretically, the ejecta is expected to emerge with a monotonic
velocity profile as a result of internal shocks that take place 
at small radii $r<10^{16}$~cm (e.g., Piran 2004).
At the end of the internal-shock stage, 
any two adjacent shells no longer collide with each other. 
Thus, we consider here only a non-increasing function of $\Gej(\tau)$; 
$\Gej^\prime(\tau) \equiv d\Gej/d\tau \leq 0$. 
Note that $\Gej$ should remain independent of 
the lab. time $t$ as each shell coasts 
without colliding with another shell. 
The corresponding velocity 
$\vej(\tau) = c\, (1-1/\Gej^2)^{1/2}$
is also non-increasing. 
The initial density of the ejecta may, however, have an arbitrary radial profile. 
The evolution of such ``stratified'' ejecta is analytically studied here, 
restricted to the unshocked ejecta.

We assume spherical symmetry. However, the calculation remains valid even if the 
explosion is driven by a jet with a small opening angle $\theta_{\rm jet}$
as long as $\Gej\gg\theta_{\rm jet}^{-1}$; the jet behaves like a 
portion of spherical ejecta since the edge of the jet is causally 
disconnected from its axis.


\subsubsection{Continuity equation of stratified ejecta}

The 4-velocity $u^{\alpha}$ for a spherically symmetric ejecta 
is written in spherical polar 
coordinates $(ct,\, r,\, \theta,\, \phi)$ as 
$u^{\alpha}=\Gej\,(c,\,\vej,\,0,\,0)$,
where $t$ indicates the lab. time, 
and $r$ is the radius measured from 
the center of the explosion. 
The continuity equation for ejecta is simply 
$\nabla_{\alpha}(\rhoej u^{\alpha})=0$. 
Here the ejecta density $\rhoej$ is measured 
in the rest frame of ejecta.  
Then the continuity equation reads 
\be
\frac{1}{c} \left.\frac{\pa}{\pa t}\right|_r (\rhoej \Gej c)+
\frac{1}{r^2} \left.\frac{\pa}{\pa r}\right|_t (r^2\, \rhoej \Gej \vej)=0,
\ee
which becomes 
\be
\label{eq:conti}
\left. \frac{\pa}{\pa t} \right|_r (\rhoej \Gej) +
\left. \frac{\pa}{\pa r} \right|_t (\rhoej \Gej \vej) +
\frac{2}{r}\,(\rhoej \Gej \vej) = 0.
\ee
We describe below a simple way 
of solving Equation (\ref{eq:conti}) 
that makes use of the Lagrangian coordinate $\tau$.

Consider a shell in ejecta that was 
ejected at time $\tau$ with velocity $\vej(\tau)$. 
The radius $r$ of this $\tau$-shell at time $t$ is given by
\be
\label{eq:radius}
r(\tau, t) = \int_{\tau}^t \, \vej(\tau)\, dt^{\prime}
= \vej(\tau)~(t-\tau).
\ee
Equation (\ref{eq:radius}) is viewed as 
a relationship among three coordinates $r$, $t$, and $\tau$. 
Any two of those can be regarded as two independent variables. 
In Equation (\ref{eq:conti}), 
the coordinates $r$ and $t$ are two independent variables. 
Since $\Gej(\tau)$ and $\vej(\tau)$ are functions of $\tau$ only, 
it is useful to have $\tau$ as 
one of the variables instead of, e.g., time $t$. 
Using relation (\ref{eq:radius}), 
we eliminate time $t$ 
from Equation (\ref{eq:conti}), 
and adopt the coordinates $\tau$ and $r$ 
as two independent variables. 
First, we partially differentiate the relation 
$r=\vej(\tau)\,(t-\tau)$ with respect to $t$ at fixed $r$, 
\be
\left.\frac{\pa r}{\pa t}\right|_r=0
=\vej^{\prime}(\tau) \left.\frac{\pa \tau}{\pa t}\right|_r \frac{r}{\vej(\tau)}+
\vej(\tau)\left(1-\left.\frac{\pa \tau}{\pa t}\right|_r \right),
\ee
which yields
\be
\left.\frac{\pa \tau}{\pa t}\right|_r=\frac{1}{1-r(\vej^{\prime}/\vej^2)},
\ee
where $\vej^{\prime}(\tau) \equiv d \vej/d\tau$.
Similarly, 
we differentiate Equation (\ref{eq:radius}) 
with respect to $r$ at fixed $t$, 
\be
\left.\frac{\pa r}{\pa r}\right|_t=1
=\vej^{\prime}(\tau) \left.\frac{\pa \tau}{\pa r}\right|_t \frac{r}{\vej(\tau)}+
\vej(\tau)\left(0-\left.\frac{\pa \tau}{\pa r}\right|_t \right),
\ee
which yields
\be
\left.\frac{\pa \tau}{\pa r}\right|_t=\frac{-1}{\vej-r(\vej^{\prime}/\vej)}.
\ee
Then we find
\beq
\label{eq:t_r}
\left.\frac{\pa}{\pa t}\right|_r 
&=& \left.\frac{\pa \tau}{\pa t}\right|_r \left.\frac{\pa}{\pa \tau}\right|_r +
    \left.\frac{\pa r}{\pa t}\right|_r \left.\frac{\pa}{\pa r}\right|_{\tau}     
 =  \frac{1}{1-r(\vej^{\prime}/\vej^2)} \left.\frac{\pa}{\pa \tau}\right|_r, \\
\label{eq:r_t}
\left.\frac{\pa}{\pa r}\right|_t 
&=& \left.\frac{\pa \tau}{\pa r}\right|_t \left.\frac{\pa}{\pa \tau}\right|_r +
    \left.\frac{\pa r}{\pa r}\right|_t \left.\frac{\pa}{\pa r}\right|_{\tau}   
 =  \frac{-1}{\vej-r(\vej^{\prime}/\vej)} \left.\frac{\pa}{\pa \tau}\right|_r +
    \left.\frac{\pa}{\pa r}\right|_{\tau}.
\eeq
We substitute Equations (\ref{eq:t_r}) and (\ref{eq:r_t}) 
into Equation (\ref{eq:conti}), and 
divide the whole equation by $\rhoej \Gej \vej$. Then we get
\be
\label{eq:conti2}
\frac{1}{\vej-r(\vej^{\prime}/\vej)} 
\left.\frac{\pa}{\pa \tau}\right|_r \left[\ln \frac{1}{\vej}\right]+
\left.\frac{\pa}{\pa r}\right|_{\tau} \left[\ln \rhoej + \ln (\Gej \vej)\right] 
+\frac{2}{r}=0,
\ee
where the term $\ln (\Gej \vej)$ vanishes 
since it is a function of $\tau$ only. 
Equation (\ref{eq:conti2}) becomes
\be
\label{eq:conti3}
\frac{-\vej^{\prime}/\vej}{\vej-r(\vej^{\prime}/\vej)} +
\left.\frac{\pa}{\pa r}\right|_{\tau} \left[\ln \rhoej \right] 
+\frac{2}{r}=0.
\ee
Finally, we integrate Equation (\ref{eq:conti3}) 
over $r$ at fixed $\tau$,
\be
\ln \left[1-r\, \frac{\vej^{\prime}}{\vej^2} \right] +
\ln \rhoej + 2\ln r = \ln \left[f(\tau)\right],
\ee
where $f(\tau)$ is an arbitrary positive function of $\tau$. 
Thus, we find an analytical solution 
of the continuity equation for stratified ejecta,
\beq
\label{eq:conti4}
\rhoej(\tau, r)
= \frac{f(\tau)}{r^2} \left[1-r\, \frac{\vej^{\prime}}{\vej^2}\right]^{-1}
= \frac{f(\tau)}{r^2} \left[1-\frac{r}{c}\, 
  \frac{\Gej^{\prime}}{(\Gej^2-1)^{3/2}}\right]^{-1}.
\eeq 
Here we have used the relation 
$\vej^{\prime}(\tau)=(c^2/\vej)(\Gej^{\prime}/\Gej^3)$.
The function $f(\tau)$ is determined below by 
the initial profile of ejecta near the center of the burst. 
The factor $1/r^2$ represents an overall side expansion 
in 3 dimensional space. 
The remaining factor inside the brackets 
is responsible for a local spread-out of the ejecta 
due to its stratification. 
For 1 dimensional plane-parallel symmetry, 
the solution $\rhoej$ remains the same as above, 
except that there is no $1/r^2$ factor.


\subsubsection{Initial profile of the ejecta}

The $\tau$-shell of energy $\delta \Eej(\tau)$, 
ejected at an ejection time $\tau$, 
coasts with its Lorentz factor $\Gej(\tau)$. 
The initial profile of the ejected flow is set by 
the central engine of the explosion. 
Note that it is completely described by two functions; 
$\Gej(\tau)$ and $\Lej(\tau) \equiv d\Eej/d\tau$. 
Here the luminosity $\Lej(\tau)$ is related to 
the mass flow rate $\dot M(\tau)$ as 
$\Lej = \Gej \dot M c^2$. 
For small radii $r$, the mass flow rate is given by
$\dot M = (4 \pi r^2 \vej)\, (\rhoej \Gej)$.
Then the initial profile of 
ejecta density $\rhoej$ at the burst place 
is expressed as
\be
\rhoej(\tau,r) = \frac{\Lej}{4 \pi r^2 \vej\, \Gej^2 c^2}.
\ee
The solution (\ref{eq:conti4}) has a limiting form $\rhoej = f(\tau)/r^2$
for small radii $r$. 
Thus we determine the function $f(\tau)$,
\be
f(\tau) = \frac{\Lej}{4 \pi \vej\, \Gej^2 c^2} 
        = \frac{\dot M}{4 \pi \vej\, \Gej}.
\ee
Hence the ejecta density is derived as
\be
\label{eq:ejecta_density}
\rhoej(\tau, r)=
\frac{\Lej(\tau)}{4 \pi r^2 \vej\, \Gej^2 c^2} 
\left[1-\frac{r}{c}\, \frac{\Gej^{\prime}}{(\Gej^2-1)^{3/2}}\right]^{-1}.
\ee
The solution (\ref{eq:ejecta_density}) is exact; 
we remark, however, that $\Gej^\prime(\tau) \leq 0$ is assumed 
in the derivation. 
For given $\Gej(\tau)$ and $\Lej(\tau)$, 
the solution (\ref{eq:ejecta_density}) allows us 
to fully understand the subsequent evolution of 
the ejected flow. 
The solution (\ref{eq:ejecta_density}) is presented in 
Uhm \& Beloborodov (2007) with no derivation.


\subsection{Jump conditions of the FS and RS} \label{section:jump_conditions}

In Section~\ref{section:shock}, 
we derived a simple form of jump conditions for shocks of arbitrary strength. 
We apply those jump conditions to the FS and RS of the blast wave, 
treating the gas in the blast as monoenergetic.

The Lorentz factors $\gamma_1$ and $\gamma_2$ are measured 
in the rest frame of the FS.  
As the ambient medium is at rest in the lab. frame, we note that 
\be
\gamma_1=\Gamma_f, 
\qquad
\gamma_2=(1-\beta \beta_f)\Gb \Gamma_f,
\qquad \mbox{and} \qquad
\gamma_{12}=\Gb.
\ee
The relative Lorentz factor $\gamma_{12}=\Gb$ 
describes the shock strength of the FS.
Then the jump conditions of the FS read
\beq
\gamma_1^2 &=& \frac{(4 \Gb^2-1)^2}{8 \Gb^2+1}
\qquad \mbox{or} \qquad
\beta_1=\frac{4\beta}{\beta^2+3}, \\
\gamma_2^2 &=& \frac{9 \Gb^2}{8 \Gb^2+1} 
\qquad \mbox{or} \qquad
\beta_2=\frac{\beta}{3}, \\
\label{eq:pf}
p_2 &=& \frac{4}{3}\, (\Gb^2-1)\, \rho_1 c^2, 
\qquad
\kappa_2=\frac{1}{3}\left(1+\frac{1}{\Gb}\right), \\
\label{eq:jump4_fs} 
\rho_2 &=& 4 \Gb\, \rho_1,
\qquad 
e_2 = 4 \Gb^2\, \rho_1 c^2.
\eeq
Thus, the Lorentz factor $\Gamma_f$ and 
the thermodynamic quantities $\rho_2$, $e_2$, $p_2$, and $\kappa_2$ 
immediately behind the FS are found 
in terms of $\Gb$ and $\rho_1$.
Here $\rho_1$ should be evaluated for the ambient medium 
immediately ahead the FS, which we denote by $\rhoFS$.

The RS is described by the same set of jump conditions 
when index 1 is replaced by 4 and index 2 by 3.
The Lorentz factors $\gamma_3$ and $\gamma_4$ are measured  
in the rest frame of the RS, 
\be
\label{eq:gamma3_4}
\gamma_3=(1-\beta \beta_r)\, \Gb \Gamma_r, 
\qquad \mbox{and} \qquad
\gamma_4=(1-\bej \beta_r)\, \Gej \Gamma_r.
\ee
The shock strength of the RS is described by 
the relative Lorentz factor $\gamma_{43}$,
\be
\gamma_{43}=(1-\beta_4 \beta_3)\, \gamma_4 \gamma_3
=(1-\beta \bej)\, \Gb \Gej.
\ee
Then the jump conditions of the RS read
\beq
\gamma_4^2 &=& \frac{(4 \gamma_{43}^2-1)^2}{8 \gamma_{43}^2+1}
\qquad \mbox{or} \qquad
\beta_4=\frac{4\beta_{43}}{\beta_{43}^2+3}, \\
\gamma_3^2 &=& \frac{9 \gamma_{43}^2}{8 \gamma_{43}^2+1} 
\qquad \mbox{or} \qquad
\beta_3=\frac{\beta_{43}}{3}, \\
\label{eq:pr}
p_3 &=& \frac{4}{3}\, (\gamma_{43}^2-1)\, \rho_4 c^2, 
\qquad
\kappa_3=\frac{1}{3}\left(1+\frac{1}{\gamma_{43}}\right), \\
\label{eq:jump4_rs} 
\rho_3 &=& 4 \gamma_{43}\, \rho_4,
\qquad 
e_3 = 4 \gamma_{43}^2\, \rho_4 c^2.
\eeq
Equation (\ref{eq:gamma3_4}) yields the Lorentz factor $\Gamma_r$, 
\be
\label{eq:Gamma_r}
\Gamma_r
=(1-\beta \beta_3)\, \Gb \gamma_3
=(1-\bej \beta_4)\, \Gej \gamma_4,
\ee
where $\gamma_3$ and $\gamma_4$ are now given 
in terms of $\g43=(1-\beta \bej)\, \Gb \Gej$ above. 
Thus, the Lorentz factor $\Gamma_r$ and 
the thermodynamic quantities $\rho_3$, $e_3$, $p_3$, and $\kappa_3$ 
immediately behind the RS are found 
in terms of $\Gb$, $\Gej$, and $\rho_4$. 
Here $\Gej$ and $\rho_4$ should be evaluated for the shell immediately ahead the RS, 
which we denote by $\GejRS$ and $\rho_4 {\rm (RS)} \equiv \rhoejRS$, respectively.

Let the RS be located at radius $\rr$ and sweep up the $\tau_r$-shell in the ejecta 
when the FS is located at radius $\rf$; 
the subscripts $r$ and $f$ refer to the RS and FS, respectively.
When three functions $\rho_1(r)$, $\Gej(\tau)$, and $\Lej(\tau)$ are known, 
we find then 
$\rhoFS=\rho_1(\rf)$, 
$\GejRS=\Gej(\tau_r)$, and 
$\rhoejRS=\rhoej(\tau_r, \rr)$; 
the ejecta density $\rhoejRS$ is given by 
the solution (\ref{eq:ejecta_density}).

Thus the Lorentz factor $\Gb$ becomes the only free parameter 
for the blast wave with known input functions $\rho_1$, $\Gej$, and $\Lej$. 
This is justified since 
we are given 6 independent equations (3 plus 3 jump conditions) 
for 7 unknowns, which are $\Gb$, $\Gamma_f$, $\Gamma_r$, 
two independent thermodynamic quantities describing the gas behind the FS, 
and another two for the gas behind the RS.


\subsection{Trajectory of the RS through ejecta}

The path of the RS needs to be consistently tracked, 
as it propagates through the ejecta. 
Consider the RS located at radius $\rr(t)$ at time $t$, 
sweeping up the $\tau_r(t)$-shell.
Equation (\ref{eq:radius}) gives the radius of 
the $\tau_r$-shell at time $t$, 
which equals $\rr(t)$; 
$\rr(t) = \vej(\tau_r[t])~(t-\tau_r[t])$.
Then we find the velocity $v_r$ of the RS,
\beq
v_r \equiv \frac{d \rr}{dt} 
&=&\vej^{\prime}(\tau_r)\, \frac{d \tau_r}{dt}\, \frac{\rr}{\vej(\tau_r)}+
\vej(\tau_r)\,\left(1-\frac{d \tau_r}{dt}\right) \\
&=&\vej(\tau_r)-\vej(\tau_r) \left[1-\rr\, 
\frac{\vej^{\prime}(\tau_r)}{\vej^2(\tau_r)} \right] \frac{d \tau_r}{dt}.
\eeq
Using the relation
\be
\frac{d \tau_r}{dt}=
\frac{d \rr}{dt}\,\frac{d \tau_r}{d \rr}=
v_r\, \frac{d \tau_r}{d \rr}, 
\ee
we find a differential equation for $d \tau_r/d \rr$,
\be
\label{eq:traj2}
\frac{d \tau_r}{d \rr} = 
\left(\frac{1}{v_r}-\frac{1}{\vej(\tau_r)}\right)\,
\left[1-\rr\, \frac{\vej^{\prime}(\tau_r)}{\vej^2(\tau_r)} \right]^{-1}.
\ee   
Equation (\ref{eq:Gamma_r}) gives the velocity $v_r$ 
in terms of $\Gb$ and $\Gej(\tau_r)$. 
Thus, Equation (\ref{eq:traj2}) allows us to follow 
the trajectory of the RS through ejecta when $\Gb$ is known; 
for an infinitesimal displacement $\delta \rr$ of the RS, 
we numerically solve Equation (\ref{eq:traj2}) to find $\delta \tau_r$.

The Lorentz factor $\Gb$ is determined below by two different methods: 
(1) customary pressure balance (see Section~\ref{section:customary_approx}) and 
(2) the mechanical model (see Section~\ref{section:mechanical_model}). 
We demonstrate that the method (1) does not satisfy 
the energy-conservation law for adiabatic blast waves.


\subsection{Adiabatic blast} \label{section:adiabatic_blast}

As the blast wave propagates through 
the ambient medium, the blast grows and 
the gas in the blast evolves hydrodynamically.   
A simple way of dealing with an adiabatic evolution 
of the blast is described here.  
Since we treat the gas as monoenergetic, 
we first study the adiabatic process of a monoenergetic gas.


\subsubsection{Adiabatic process of a monoenergetic gas}

Consider a relativistic monoenergetic gas, which has 
the pressure $p$, energy density $e$, mean Lorentz factor $\bar \gamma$, 
volume $V$, number density $n$, and particle number $N$. 
An adiabatic process of the gas is defined by $d(eV)= -p\,dV$. 
When the particle number is conserved, we have 
$N=nV=\mbox{const}$, or $dV=-N\,dn/n^2$.
Recalling Equation (\ref{eq:pe_mono}) 
for $p$ and $e$ of the monoenergetic gas, we find 
\be
d(nV\, \bar \gamma m c^2) = 
n\, mc^2\, \frac{{\bar \gamma}^2-1}{3 \bar \gamma} \left[\frac{N}{n^2}\,dn \right],
\ee
which yields
\be
\label{eq:mono_adiab}
\frac{3 \bar \gamma}{(\bar \gamma^2 -1)}\, d \bar \gamma = 
\frac{1}{n}\, dn.
\ee
Here $m$ denotes the particle mass in the gas.
We integrate Equation (\ref{eq:mono_adiab}) to find 
\be
\label{eq:adiab}
n \propto (\bar \gamma^2 -1)^{3/2}
\qquad 
\mbox{and}
\qquad
p \propto \frac{1}{\bar \gamma} (\bar \gamma^2 -1)^{5/2}.
\ee
Note that $p \propto \bar \gamma^4$ is verified 
for the adiabatic process of an ultra-relativistic 
gas $\bar \gamma \gg 1$. 
Using Equation (\ref{eq:mono}), i.e.,
$\kappa = \frac{1}{3}(1+1/\bar \gamma)$,
or $\bar \gamma = 1/(3 \kappa -1)$, 
we re-write relation (\ref{eq:adiab}) as
\be
\label{eq:adiab2}
p \propto \frac{\kappa^{5/2}~(\frac{2}{3}-\kappa)^{5/2}}
{(\kappa-\frac{1}{3})^4} 
\equiv p_m(\kappa).
\ee
Here we have defined the function $p_m(\kappa)$ 
for the right-hand side. 
The function $p_m(\kappa)$ is monotonically decreasing in its valid range, 
$\frac{1}{3} < \kappa < \frac{2}{3}$.
The proportionality constant in Equation (\ref{eq:adiab2}) 
is related to the entropy of the gas, 
which is conserved for the adiabatic process.


\subsubsection{Evolution of adiabatic blast} \label{section:blast_evolution}

We discretize the external ambient medium and ejecta 
into spherical mass shells $\delta m$, 
and use a Lagrangian description for the blast wave. 
Each $\delta m$ is impulsively heated at some point 
by a shock front (FS or RS), 
acquiring its initial pressure $p$ and quantity $\kappa$, 
which are given by the jump conditions 
(see Section~\ref{section:jump_conditions}).
Using Equation (\ref{eq:adiab2}), 
we can track the subsequent adiabatic evolution 
of each $\delta m$ if we know the evolution of its pressure; 
the initial $p$ and $\kappa$ determine the proportionality constant, 
and we find numerically the new quantity $\kappa$ 
when $\delta m$ is at new pressure $p$. 
All other thermodynamic quantities of $\delta m$ 
can then be found accordingly. 
For instance, the volume $\delta V$ 
of the mass shell is obtained as
\be
\label{eq:deltaV}
\delta V 
= \frac{m c^2}{p}\, \frac{(\bar \gamma^2-1)}{3 \bar \gamma}\, \delta N
= \frac{m c^2}{p}\, \frac{\kappa (\frac{2}{3}-\kappa)}{(\kappa-\frac{1}{3})}\, \delta N,
\ee
using Equation (\ref{eq:pe_mono}) and $\bar \gamma=1/(3\kappa-1)$. 
This volume $\delta V$ is defined 
in the rest frame of the mass shell.
The particle number $\delta N$ of the mass shell is calculated
when $\delta m$ is shocked by a shock front, 
by making use of the particle number flux 
in Equation (\ref{eq:mass_flux}).

\subsubsection{Energy of the blast} \label{section:blast_energy}

We may then calculate the energy of the blast. 
When the blast has an instantaneous Lorentz factor $\Gb$, 
the total energy of the entire blast is evaluated 
in the lab. frame by integrating the $00$-component of 
energy-momentum tensor over the volume 
of each $\delta m$:
\beq
E_{\rm blast} 
= \sum_{\lbrace \delta m \rbrace} 
\left[\Gb^2\,(e+p)-p \right] \left(\frac{\delta V}{\Gb} \right) 
= \sum_{\lbrace \delta m \rbrace} 
\left[\Gb\, e\,\delta V +
\left(\Gb - \frac{1}{\Gb}\right)\, p\, \delta V \right].                  
\eeq
Due to the Lorentz contraction, 
$\delta V/\Gb$ is the volume of $\delta m$
in the lab. frame. 
Replacing the energy $e\, \delta V$ by 
$(\bar \gamma\, m c^2)\,\delta N$ 
and using Equation (\ref{eq:deltaV}) for $p\, \delta V$, 
we find 
\be
\label{eq:Eblast}
E_{\rm blast}= 
\sum_{\lbrace \delta m \rbrace} 
\left[\Gb\, \bar \gamma + \frac{1}{3} 
\left(\Gb-\frac{1}{\Gb}\right) \left(\bar \gamma-\frac{1}{\bar \gamma}\right)\right] 
\left(m c^2\, \delta N \right).
\ee
We emphasize that the second term here needs to be included 
in order to correctly express the energy of blast. 
For relativistic blast waves, $\Gb^2 \gg 1$, 
Equation (\ref{eq:Eblast}) becomes
\beq
E_{\rm blast}               
\simeq \sum_{\lbrace \delta m \rbrace} 
\Gb \left[\frac{1}{3} \left(4 \bar \gamma-\frac{1}{\bar \gamma}\right) \right]
\left(m c^2\, \delta N \right) 
= \sum_{\lbrace \delta m \rbrace} 
\Gb \left[ \frac{(1-\kappa)(3\kappa+1)}{(3\kappa-1)} \right]
\left(m c^2\, \delta N \right).        
\eeq

The energy $E_4$ of unshocked ejecta (region 4) is 
easily found as $E_4=\int_{\tau_r}^{\infty} \Lej(\tau)\,d\tau$,
where $\tau_r$ indicates the location of the RS in the ejecta. 
The energy of region 1 is negligible 
since the ambient medium is at rest in the lab. frame. 
Thus, the total energy of the entire system is 
obtained as $E_{\rm tot}=E_{\rm blast}+E_4$.


\subsection{Customary pressure balance: $\pf=\pr$} \label{section:customary_approx}

A customary approximation assumes a pressure balance across the blast wave; 
i.e., the pressure $\pf$ at the FS is equated to the pressure $\pr$ at the RS.  
Equations (\ref{eq:pf}) and (\ref{eq:pr}) give 
the pressures $\pf=p_2$ and $\pr=p_3$, 
\be
\pf=\frac{4}{3}\,(\Gb^2-1)\, \rho_1 c^2, 
\qquad
\pr=\frac{4}{3}\, (\g43^2-1)\, \rho_4 c^2.
\ee 
For relativistic blast waves ($\Gej\gg 1$ and $\Gb\gg 1$), 
the pressure balance $\pf=\pr$ determines 
the instantaneous $\Gb$ of the blast wave (Beloborodov \& Uhm 2006), 
\be 
\label{eq:custom}
\Gb=\Gej\left[1+2\, \Gej\left(\frac{\rho_1}{\rhoej}\right)^{1/2}\right]^{-1/2}, 
\qquad \pf=\pr,
\ee
denoting $\rho_4=\rhoej$. 
Recall that $\rho_1=\rhoFS$, $\rhoej=\rhoejRS$, and $\Gej=\GejRS$ 
should be used here; see Section \ref{section:jump_conditions}.

The solution (\ref{eq:custom}) indicates that 
the dynamical evolution of the blast wave is determined by purely 
input parameters $\rho_1$, $\rhoej$, and $\Gej$ of regions 1 and 4. 
The solution $\Gb$ has no information on the thermodynamical status 
of the gas in blast. 
This observation makes us to doubt 
the validity of the solution (\ref{eq:custom}). 
The assumption $\pf=\pr$ itself is then doubted. 
As we demonstrate below, 
the solution (\ref{eq:custom}) in fact does {\it not} satisfy 
the energy conservation for adiabatic blast waves.


\subsubsection{Example model} \label{section:example_model}

The initial setup of an explosion is specified by 
three functions $\rho_1$, $\Gej$, and $\Lej$, which can be 
arbitrary as long as $\Gej^\prime(\tau) \leq 0$. 
We consider a simple example model that assumes 
\beq
\label{eq:example_model}
\Lej(\tau)=L_0=10^{52}~\mbox{erg/s},
\qquad
\Gej(\tau)=500-9\tau,
\qquad
0 \leq \tau \leq \tau_b=50~\mbox{s}. 
\eeq
The luminosity $\Lej$ remains at a constant $L_0$ 
during the duration $\tau_b$ of the burst. 
The Lorentz factor $\Gej$ decreases linearly from 500 to 50.
The total energy $E_b$ of the burst is simply $E_b = L_0\,\tau_b$. 
The ambient medium is assumed to have a uniform 
density $n_1=\rho_1/m_p=1~\mbox{cm}^{-3}$. 
Here $m_p$ is the proton mass.

\begin{figure}
\begin{center}
\includegraphics[width=15cm]{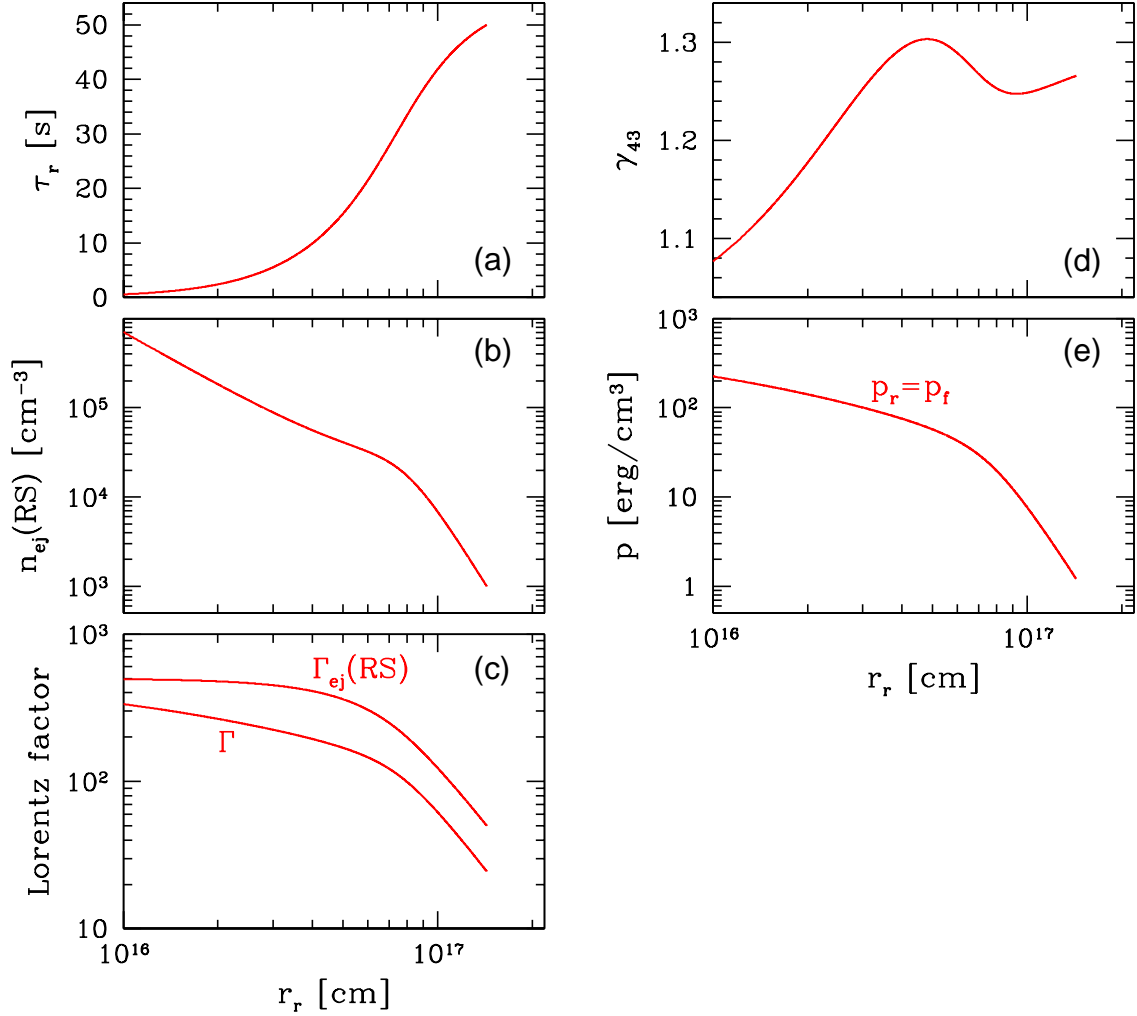}
\caption{
Numerical solution for the blast-wave driven by 
the example burst specified in Equation (\ref{eq:example_model}); 
$\Lej(\tau)=L_0=10^{52}~\mbox{erg/s}$ and $\Gej(\tau)=500-9\tau$ 
for $0 \leq \tau \leq \tau_b=50~\mbox{s}$. 
The ambient medium density is assumed to be $n_1=\rho_1/m_p=1~\mbox{cm}^{-3}$. 
Here $m_p$ is the proton mass. 
This solution is found as described in Section \ref{section:example_model}, 
using Equation (\ref{eq:custom}) of the pressure balance $\pr=\pf$. 
Panel (a) shows the $\tau_r$-shell passing through the RS at radius $\rr$. 
Panel (b) is the ejecta density $\nejRS = \rhoejRS/m_p$ of the $\tau_r$-shell.  
Panel (c) shows the Lorentz factor $\GejRS$ of the $\tau_r$-shell and $\Gb$ of the blast, 
which yields the relative Lorentz factor $\g43$ in panel (d).
Panel (e) shows that a pressure balance $p=\pr=\pf$ is assumed across the blast. 
However, this numerical solution does not satisfy the energy-conservation law
for the adiabatic blast wave; see Figure \ref{fig:etot_custom}.
} 
\label{fig:dyn_custom}
\end{center}      
\end{figure}

For this example burst, we find the evolution of the blast wave as follows. 
Suppose that, at time $t$, 
the RS is located at radius $\rr(t)$, 
the FS is located at radius $\rf(t)$,
the $\tau_r(t)$-shell passes through the RS, and 
the blast has the Lorentz factor $\Gb(t)$. 
We evaluate 
$\rhoFS=\rho_1(\rf)$, 
$\GejRS=\Gej(\tau_r)$, and 
$\rhoejRS=\rhoej(\tau_r, \rr)$ (Equation (\ref{eq:ejecta_density})), 
which in turn gives $\g43$, $\gamma_3$, 
$\Gamma_r$ (Equation (\ref{eq:Gamma_r})), and $\Gamma_f$. 
For an infinitesimal displacement $\delta \rr$ of the RS, 
we numerically solve Equation (\ref{eq:traj2}) to find $\delta \tau_r$.
Equation (\ref{eq:radius}), $r=\vej(\tau)\,(t-\tau)$, 
i.e., $t=r/\vej(\tau)+\tau$, 
gives then the time for the new location of the RS 
($\rr + \delta \rr$ in radius and $\tau_r + \delta \tau_r$ in ejecta). 
Thus, we find the time interval $\delta t$ for this displacement $\delta \rr$. 
The new location of the FS is found by its displacement $\delta \rf$ during $\delta t$ 
with the velocity given by its Lorentz factor $\Gamma_f$. 
We evaluate 
$\rhoFS$, $\GejRS$, and $\rhoejRS$ again for the new location. 
Then the solution (\ref{eq:custom}) determines the Lorentz factor $\Gb$ 
of the blast for the new location. 
The result is shown in Figure~\ref{fig:dyn_custom}.

The solution (\ref{eq:custom}) also determines the pressure $p=\pf=\pr$ across the blast.
This enables us to track the adiabatic evolution of the mass shells 
in blast (see Section~\ref{section:blast_evolution}), 
and to find the total energy $E_{\rm tot}$ of the entire system 
(see Section~\ref{section:blast_energy}). 
In Figure~\ref{fig:etot_custom}, we show the resulting total energy $E_{\rm tot}$. 
Apparently, the energy conservation is not satisfied; 
$E_{\rm tot}$ has decreased by a factor of 5 by the moment 
the RS crosses the last shell ($\tau = 50$ s) in the ejecta.

\subsubsection{What is wrong?}

The spherical expansion of ejecta was completely described by 
the solution (\ref{eq:ejecta_density}). 
The propagation of the RS through ejecta was found 
self-consistently by Equation (\ref{eq:traj2}). 
The conservation laws of energy-momentum tensor and mass flux 
were explicitly applied across both the FS and the RS (the jump conditions). 
The gas in blast, however, was not required to obey those conservation laws; 
i.e., the pressure balance $\pf=\pr$ omits
the physics laws that should govern the gas in blast. 
Evidently, this is why the energy conservation 
for the adiabatic blast wave was not satisfied above. 
Note that one among three independent conservation laws was 
effectively applied to the blast, 
since we tracked the adiabatic evolution of the mass shells in blast 
in order to find the total energy of the system.

Adiabatic expansion of the gas in blast implies a $pdV$ work done to the gas itself. 
This work needs to be converted to the kinetic energy of the bulk motion of blast. 
Clearly, such a conversion mechanism is absent from the solution (\ref{eq:custom}) 
as it depends only on input parameters of regions 1 and 4.
A correct modeling for the blast wave should indicate 
a mechanism that the dynamical variable $\Gb$ is affected 
by the thermodynamic status of the gas in blast.

Applying the conservation laws of energy-momentum tensor 
and mass flux to everywhere on the blast,  
we developed a simple ``mechanical model'' for the blast wave 
(Beloborodov \& Uhm 2006). 
The mechanical model successfully resolves
the energy-violation problem, because it replaces 
the pressure balance $\pf=\pr$ by the physics laws. 
We summarize the model below including more detailed equations.

%
%

\section{Mechanical model for relativistic blast waves} \label{section:mechanical_model}

The gas in the blast wave flows radially with the 4-velocity
$u^\alpha=\gamma\,(1,\, \beta,\, 0,\, 0)$ in spherical coordinates 
$(ct,\, r,\, \theta,\, \phi)$, 
where the metric $ds^2=-c^2 dt^2+dr^2+r^2 d\theta^2+r^2 \sin^2 \theta d\phi^2$ has 
the determinant $g=-r^4 \sin^2 \theta$. 
For any scalar function $f$, the covariant divergence of 
$f u^{\alpha}$ is given as (e.g., see Carroll 2004),
\beq
\nabla_{\alpha}(f u^{\alpha})
&=&\frac{1}{\sqrt{-g}}\, \pa_{\alpha}\left(\sqrt{-g}\, f u^{\alpha}\right) 
= \frac{1}{r^2}\, \pa_{\alpha}(r^2 f u^{\alpha}) \\
\label{eq:divergence}
&=&\frac{1}{r^2 c}\, \frac{d}{dt}(r^2 f \gamma)+f \gamma\, \frac{\pa \beta}{\pa r},
\eeq
where $\frac{d}{dt}\equiv \frac{\pa}{\pa t} 
+c\beta\frac{\pa}{\pa r}$ is the convective derivative.
The rest-mass conservation 
$\nabla_{\alpha}(\rho u^{\alpha})=0$ reads then
\be
\label{eq:rho}
  \frac{1}{r^2 c}\frac{d}{dt}\left(r^2\rho\gamma\right)
   +\rho\gamma\,\frac{\partial \beta}{\partial r}=0.
\ee
The energy-momentum tensor for a perfect fluid is written as
\be
T^{\alpha \,\mu}=h\,u^{\alpha}u^{\mu}+g^{\alpha \,\mu}\,p,
\qquad h \equiv e+p.
\ee
The conservation 
$\nabla_\mu T_\alpha^{\;\mu}
=\nabla_\mu (h\,u_\alpha u^\mu+\delta_\alpha^{\;\mu}\,p)=0$ 
gives two independent equations ($\alpha=0,1$)
\be
\label{eq:ua}
  \frac{1}{r^2 c}\frac{d}{dt}\left(r^2h\gamma u_\alpha\right) 
   +h\gamma u_\alpha\frac{\partial\beta}{\partial r}+\partial_\alpha p=0,
\ee 
where Equation (\ref{eq:divergence}) is used. 
For $\alpha=1$, Equation (\ref{eq:ua}) becomes 
\be
\label{eq:alpha1}
\frac{1}{r^2 c}\frac{d}{dt}\left(r^2h\gamma^2\beta\right)
= -\frac{\partial p}{\partial r}
  -h\gamma^2\beta\frac{\partial\beta}{\partial r}. 
\ee          
Instead of $\nabla_\mu T_0^{\;\mu}=0\; (\alpha=0)$, 
we use the projection
$u^\alpha \nabla_\mu T_{\alpha}^{\;\mu}=0$. 
Since $u^\alpha \, u_\alpha=-1$ and $u^\alpha \nabla_\mu u_\alpha=0$, 
the projection becomes 
\be
\nabla_\mu (h\,u^\mu)=u^\alpha \nabla_\alpha\, p,
\ee 
which yields   
\be
\label{eq:cons}
\frac{1}{r^2 c}\frac{d}{dt}\left(r^2h\gamma\right)
=\frac{\gamma}{c}\,\frac{d p}{d t}-h\gamma\,\frac{\partial\beta}{\partial r}.
\ee
We apply three independent equations~(\ref{eq:rho}), 
(\ref{eq:alpha1}), and (\ref{eq:cons}) to 
the gas between the FS and the RS, and make the approximation 
\be
  \gamma(t,r)=\Gb(t), \qquad \partial\beta/\partial r=0, \qquad \rr<r<\rf,
\ee
where $\rr(t)$ and $\rf(t)$ are the instantaneous radii of the RS and FS, 
respectively. 
Then the integration of three equations~(\ref{eq:rho}), 
(\ref{eq:alpha1}), and (\ref{eq:cons}) over $r$ 
between $\rr$ and $\rf$ (at $t=\mbox{const}$) yields 
\beq
\label{eq:mech1}
&&\frac{1}{r^2c} \frac{d}{dt}\left(r^2\Sigma\, \Gb \right)-
  \Gb\left[\rhor(\beta-\beta_r)+\rhof(\beta_f-\beta) \right]=0, \\
\label{eq:mech2}  
&&\frac{1}{r^2c} \frac{d}{dt}\left(r^2 H \Gb^2\beta \right)-
  \Gb^2\beta \left[\hr(\beta-\beta_r)+\hf(\beta_f-\beta) \right]=\pr-\pf, \\
&&\frac{1}{r^2c} \frac{d}{dt}\left(r^2 H \Gb \right)-
  \Gb\left[\hr(\beta-\beta_r)+\hf(\beta_f-\beta) \right]=
  \frac{\Gb}{c} \frac{d}{dt}P- \nonumber \\
\label{eq:mech3}  
&&\Gb\left[\pr(\beta-\beta_r)+\pf(\beta_f-\beta) \right],
\eeq
where $\Sigma\equiv\int_{\rr}^{\rf}\rho\,dr$, 
$H\equiv\int_{\rr}^{\rf} h\,dr$, 
$P\equiv\int_{\rr}^{\rf} p\,dr$, 
$c\beta_r=d\rr/dt$, and $c\beta_f=d\rf/dt$. 
Here we have used an identity for any function $f(t,r)$,
\beq
&&\int_{\rr(t)}^{\rf(t)}\left[\frac{d}{dt}f(t,r) \right]\,dr=
  \frac{d}{dt}\left[\int_{\rr(t)}^{\rf(t)}f(t,r)\,dr \right]- \nonumber \\
&&c\left[\fr(t)\{\beta(t)-\beta_r(t)\}+\ff(t)\{\beta_f(t)-\beta(t)\} \right],
\eeq
where $\fr(t)\equiv f(t,\rr(t))$ and $\ff(t)\equiv f(t,\rf(t))$.
The relativistic blast is a very thin shell, $\rf-\rr\sim r/\Gb^2\ll r$,
so we used $\rf\approx\rr\approx r$ when calculating the integrals.

We simplify Equations~(\ref{eq:mech1})-(\ref{eq:mech3}) 
by making use of $\Gb \gg 1$. 
The jump conditions at the FS give $\Gamma_f^2=2\Gb^2$, 
$\beta_f-\beta=1/(4\Gb^2)$, $e_f=3\pf$, 
and $\hf=e_f+\pf=4\pf \gg \rhof\, c^2$. 
The convective derivative $d/dt=c\beta\, d/dr$ may be 
replaced by $c\, d/dr$ everywhere, 
and $\Gb^2\beta$ by $\Gb^2$ in Equation~(\ref{eq:mech2}). 
Then we get 
\beq
\label{eq:1}
   \frac{\Gb}{r^2}\frac{d}{dr}\left(r^2\;\Sigma\;\Gb\right) & = & 
  \rhor(\bb-\br)\Gb^2+\frac{1}{4}\,\rhof, \\
\label{eq:2}
   \frac{1}{r^2}\frac{d}{dr}\left(r^2H\Gb^2\right)
    & = & \hr (\bb-\br)\Gb^2+\pr,\\
\label{eq:3}
   \frac{\Gb}{r^2}\frac{d}{dr}\left(r^2H\;\Gb\right)
        & = & \Gb^2\frac{dP}{dr}+(\hr-\pr)(\bb-\br)\Gb^2+\frac{3}{4}\,\pf.
\eeq 
The FS jump conditions give 
\be
\rhof = 4 \Gb\, \rho_1,
\qquad
\pf = \frac{4}{3}\, \Gb^2\, \rho_1 c^2.
\ee
Recalling $\beta_3=(\beta-\beta_r)/(1-\beta \beta_r)$ and 
the RS jump condition $\beta_3=\beta_{43}/3$, we find 
\be
\beta-\beta_r=
\frac{\beta_{43}}{\Gb^2\,(3-\beta \beta_{43})}=
\frac{\beta_{43}}{\Gb^2\,(3-\beta_{43})},
\ee
where the second equality is valid for $\g43 \ll \Gb$.
For the relativistic blast wave, 
the Lorentz factor $\g43=(1-\beta \bej)\, \Gb \Gej$ becomes
\be
\label{eq:b43} 
  \gamma_{43}=\frac{1}{2} \left(\frac{\Gej}{\Gb}+\frac{\Gb}{\Gej}\right),
  \qquad \beta_{43}=\frac{\Gej^2-\Gb^2}{\Gej^2+\Gb^2}.
\ee
Then we find  
\be
\bb-\br=\frac{\Gej^2-\Gb^2}{2\Gb^2(\Gej^2+2\Gb^2)}.
\ee
The jump conditions at the RS are 
\be
\rhor = 4 \gamma_{43}\, \rhoej,
\qquad 
e_r = 4 \gamma_{43}^2\, \rhoej c^2,
\qquad
\pr = \frac{4}{3}\, (\gamma_{43}^2-1)\, \rhoej c^2,
\ee
which yield 
\beq
\label{eq:RS}
\rhor &=& 2\left(\frac{\Gej}{\Gb}+\frac{\Gb}{\Gej}\right)\rhoej,  
\qquad
\pr=\frac{1}{3}\left(\frac{\Gej}{\Gb}-\frac{\Gb}{\Gej}\right)^2\rhoej c^2, \\   
\hr &=& \frac{4}{3}\left(\frac{\Gej^2}{\Gamma^2}+\frac{\Gamma^2}{\Gej^2}+1\right)
     \rhoej c^2.
\eeq
Here Equation~(\ref{eq:b43}) has been used.

This leaves four unknowns in Equations~(\ref{eq:1})-(\ref{eq:3}): 
$\Sigma$, $H$, $P$, and $\Gb$. One more equation is required to close
the set of coupled differential equations. 
We propose the following approximate relation:
\be
\label{eq:4P}
   H-\Sigma c^2=4P.
\ee  
As shown in Beloborodov \& Uhm (2006), it is accurate in the limits of 
both an ultra-relativistic RS and a non-relativistic RS, 
and should be a reasonable approximation in 
an intermediate case.

%
%

\section{Discussion}

A dynamical evolution of the blast wave is found for the mechanical model as follows.
For an infinitesimal displacement of the blast, 
we numerically solve the coupled equations (\ref{eq:1})-(\ref{eq:3}) 
and (\ref{eq:4P}) of the mechanical model, 
and find the instantaneous Lorentz factor $\Gb$ and 
integrated quantities $H$, $\Sigma$, and $P$.
We also solve the differential equation (\ref{eq:traj2}) 
to get the RS trajectory through the ejecta.

\begin{figure}
\begin{center}
\includegraphics[width=11cm]{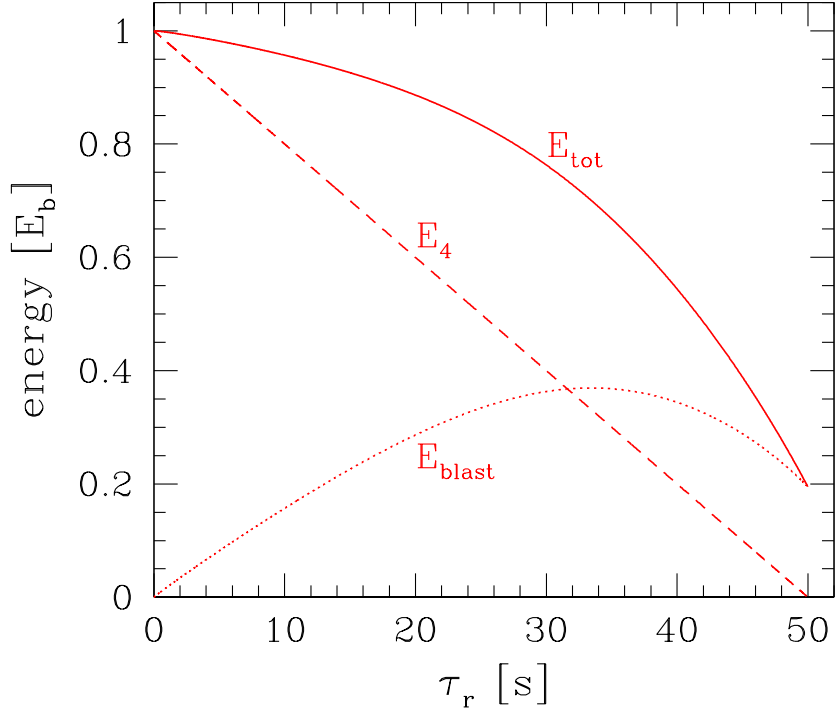}
\caption{
Energy $E_{\rm blast}$ of the blast (dotted line), 
the energy $E_4$ of region 4 (dashed line), and 
the total energy $E_{\rm tot}$ of the entire system 
(solid line; $E_{\rm tot}=E_{\rm blast}+E_4$) 
for the blast-wave evolution shown in Figure \ref{fig:dyn_custom}. 
The dynamical evolution in Figure \ref{fig:dyn_custom} 
is found by using the pressure balance $\pr=\pf$ 
for the example burst specified in Equation (\ref{eq:example_model}); 
$\Lej(\tau)=L_0=10^{52}~\mbox{erg/s}$ and $\Gej(\tau)=500-9\tau$ 
for $0 \leq \tau \leq \tau_b=50~\mbox{s}$. 
The ambient medium density is assumed to be $n_1=\rho_1/m_p=1~\mbox{cm}^{-3}$. 
The energy $E_b$ here is the total energy ejected by the burst; 
$E_b=L_0 \tau_b=5 \times 10^{53}~\mbox{erg}$. 
The energies $E_{\rm blast}$, $E_4$, and $E_{\rm tot}$ are shown in the units of $E_b$ 
for the $\tau_r$-shell, i.e., the location of the RS in the ejecta. 
We precisely track the adiabatic evolution of the mass shells 
in the blast (see Section~\ref{section:blast_evolution}) 
and find the total energy $E_{\rm tot}$ of the entire system 
(see Section~\ref{section:blast_energy}). 
However, the resulting total energy is clearly not conserved above; 
it has decreased by a factor of 5 
by the moment the RS crosses the last shell ($\tau=50~\mbox{s}$) in the ejecta.
This demonstrates that 
the solution (\ref{eq:custom}) derived from the 
pressure balance $\pf=\pr$ violates the energy-conservation law
significantly for the adiabatic blast wave.  
} 
\label{fig:etot_custom}
\end{center}      
\end{figure}

\begin{figure}
\begin{center}
\includegraphics[width=15cm]{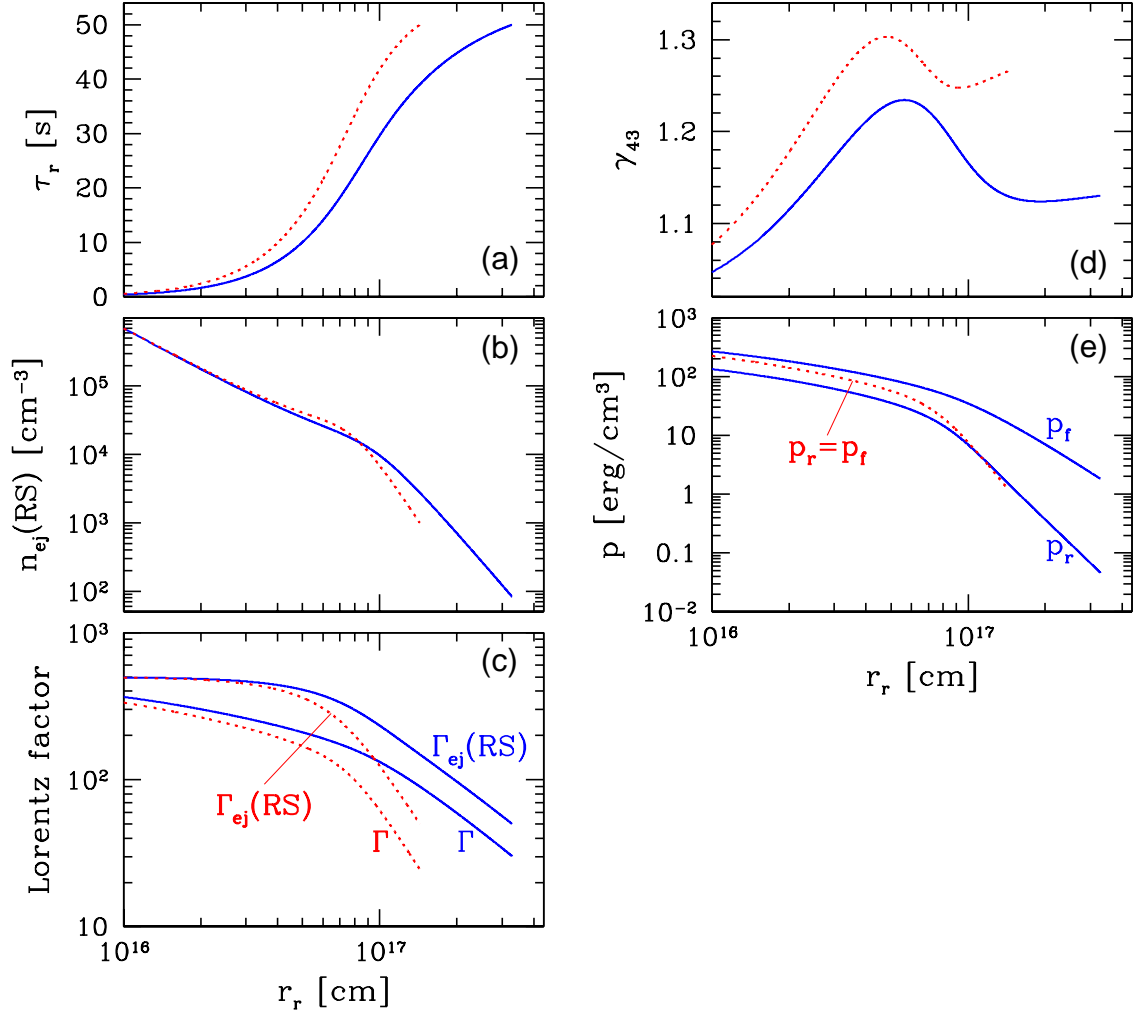}
\caption{
Numerical solutions for the blast-wave driven by the same example burst 
described in Equation (\ref{eq:example_model}). 
The ambient medium density is also the same; $n_1=\rho_1/m_p=1~\mbox{cm}^{-3}$.  
The solid (blue) curves are calculated using the mechanical model 
(see Section~\ref{section:mechanical_model}). 
The dotted (red) curves show, for comparison, 
the solution of Figure \ref{fig:dyn_custom} 
(found for the pressure balance). 
Two sets of solutions differ significantly; in particular, 
the blast wave found for the mechanical model decelerates slower 
and propagates farther (Panel c) 
until the RS arrives at the same last shell ($\tau = 50$ s) in the ejecta (Panel a). 
The solid curves satisfy the energy-conservation law for the adiabatic blast wave; 
see Figure~\ref{fig:etot}.
} 
\label{fig:dyn}
\end{center}      
\end{figure}

For the same example model as used for the customary pressure balance $\pr=\pf$,
we find the blast-wave evolution. 
Recall that the example burst is described in Equation (\ref{eq:example_model}) 
and the ambient medium is assumed to have the density $n_1=\rho_1/m_p=1~\mbox{cm}^{-3}$. 
The result found for the mechanical model is shown in Figure~\ref{fig:dyn} (in solid lines); 
for comparison, the solution found for the pressure balance 
is shown together (in dotted lines). 
Note that two sets of solutions differ significantly; in particular, 
the blast wave obtained for the mechanical model decelerates slower 
and propagates farther (see Panel c) 
until the RS arrives at the same last shell ($\tau = 50$ s) in the ejecta (see Panel a). 
For the pressure balance, on the other hand, 
the energy loss shown in Figure~\ref{fig:etot_custom} is 
responsible for the earlier deceleration of the blast wave.

\begin{figure}
\begin{center}
\includegraphics[width=11cm]{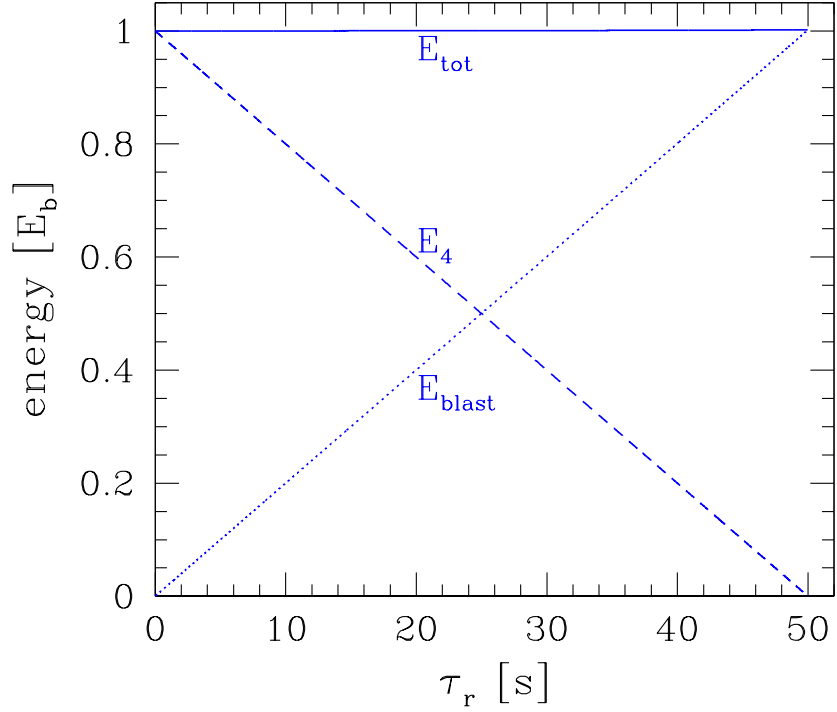}
\caption{
Energy $E_{\rm blast}$ of the blast (dotted line), 
the energy $E_4$ of region 4 (dashed line), and 
the total energy $E_{\rm tot}$ of the entire system 
(solid line; $E_{\rm tot}=E_{\rm blast}+E_4$) 
for the blast-wave evolution shown in Figure~\ref{fig:dyn} 
(i.e., the solid blue curves found for the mechanical model). 
The energy $E_b$ is the same as in Figure~\ref{fig:etot_custom}.  
The energies $E_{\rm blast}$, $E_4$, and $E_{\rm tot}$ are shown in the units of $E_b$ 
for the $\tau_r$-shell, i.e., the location of the RS in the ejecta. 
The blast energy $E_{\rm blast}$ here is found 
as in Equation (\ref{eq:Eblast_mech}) for the mechanical model. 
The total energy is precisely conserved above; 
thus, the mechanical model successfully resolves the energy-violation problem 
seen in Figure~\ref{fig:etot_custom}.
} 
\label{fig:etot}
\end{center}      
\end{figure}

The energy of the blast is easily found for the mechanical model;  
it is evaluated in the lab. frame by integrating the $00$-component of 
energy-momentum tensor over volume, 
\be
\label{eq:Eblast_mech}
E_{\rm blast} 
= \int_{\rr}^{\rf} \left[\Gb^2\, (e+p)-p \right] \left(4 \pi r^2 dr \right)
\simeq 4 \pi r^2\,(\Gb^2 H - P).
\ee
We find this blast energy for the same example model above 
and show the result in Figure~\ref{fig:etot}. 
The total energy $E_{\rm tot}$ of the entire system is precisely conserved; 
the mechanical model is indeed a successful remedy for the energy-violation problem.

The energy of the blast can alternatively be found 
by using expression (\ref{eq:Eblast}).
In order to track the adiabatic evolution of the mass shells 
in blast (see Section~\ref{section:blast_evolution}) 
and to find the blast energy, 
we need to know a pressure profile for the mass shells.
In case of the mechanical model, 
an instantaneous pressure profile for the blast may be 
approximated by a quadratic function of $r$ ($\rr <r< \rf$), which 
(1) matches two boundary values ($\pr$ at $\rr$ and $\pf$ at $\rf$) 
and (2) satisfies the integrated pressure $P$. 
The boundary values are met by a quadratic function $p(r)=a(r-b)^2+c$ with 
\beq
b &=& \frac{1}{2}\, (\rf+\rr)-
      \frac{1}{2a}\; \frac{\pf-\pr}{\rf-\rr}, \\
c &=& \frac{1}{2}\, (\pf+\pr)-\frac{a}{4}\, (\rf-\rr)^2-
      \frac{1}{4a}\, \left(\frac{\pf-\pr}{\rf-\rr} \right)^2.
\eeq
The remaining unknown $a$ is determined such that 
$P=\int_{\rr}^{\rf} p(r)\, dr$. 
An upper bound for $a$ is required since $p(r)$ is positive 
everywhere on the blast; 
$a < (\sqrt{\pf}+\sqrt{\pr})^2/(\rf-\rr)^2$.

\begin{figure}
\begin{center}
\includegraphics[width=11cm]{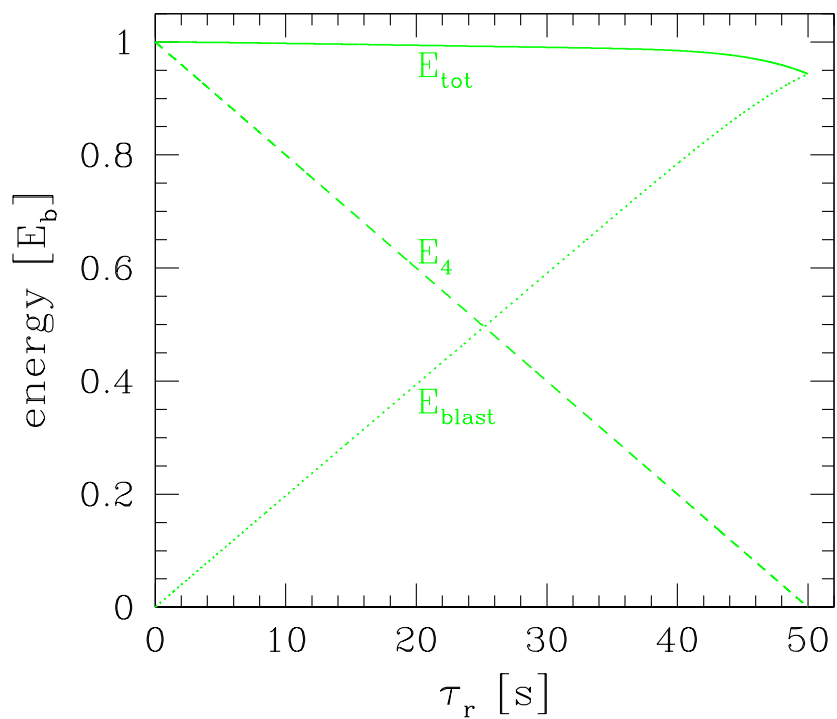}
\caption{
Same as in Figure~\ref{fig:etot}, 
except for an alternative method 
of finding the blast energy $E_{\rm blast}$. 
The blast energy here is found 
by evaluating expression (\ref{eq:Eblast}), 
while making use of an approximate pressure profile 
of a quadratic function for the blast 
(see Section 5; 4th paragraph).
The total energy $E_{\rm tot}$ is conserved within about 5 \% above; 
thus, the quadratic pressure profile is a reasonably good 
approximation for the mechanical model. 
} 
\label{fig:etot_quad}
\end{center}      
\end{figure}

We find this quadratic profile for the dynamical evolution 
shown in Figure~\ref{fig:dyn} 
(i.e., the solid lines found for the mechanical model) 
and evaluate expression (\ref{eq:Eblast}) to find 
the blast energy. 
The result is shown in Figure~\ref{fig:etot_quad}. 
The total energy $E_{\rm tot}$ is conserved within about 5 \% 
for this example. 
Thus, the quadratic pressure profile should be a reasonably good 
approximation for the mechanical model.

\begin{figure}
\begin{center}
\includegraphics[width=11cm]{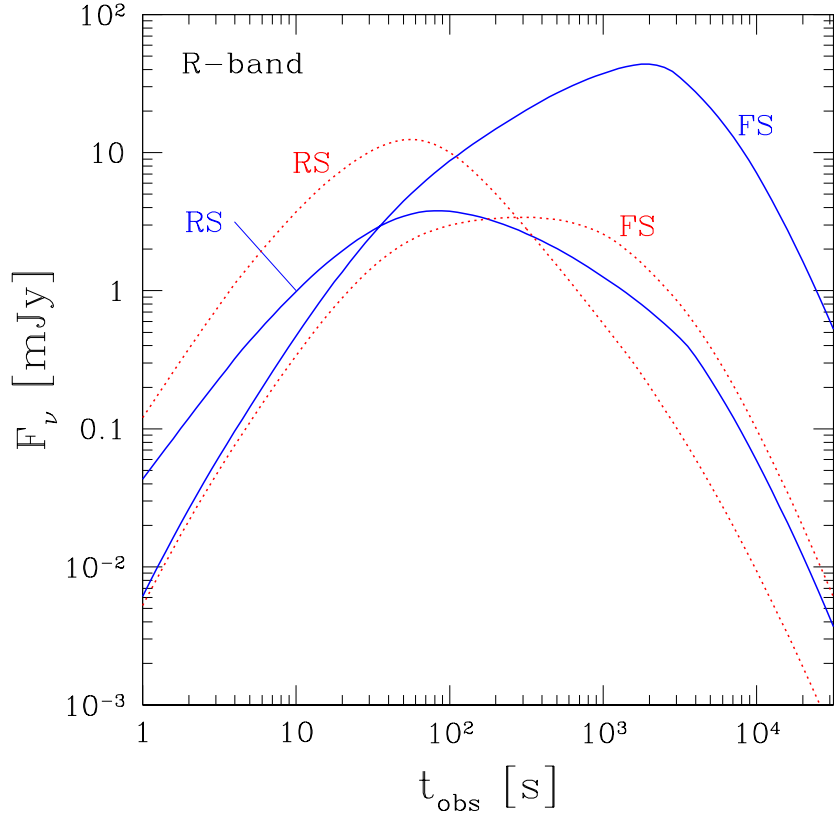}
\caption{
Afterglow light curves in $R$ band, obtained for the same example burst 
as in Figure~\ref{fig:dyn}. 
FS and RS indicate the emissions from region 2 (FS-shocked) 
and 3 (RS-shocked), respectively. 
The solid (blue) curves are calculated using the mechanical model,
corresponding to the solid (blue) curves in Figure~\ref{fig:dyn}.
The dotted (red) curves are obtained using the pressure balance, 
corresponding to the dotted (red) curves in Figure~\ref{fig:dyn}.
The emission parameters are 
$\epsilon_B=0.01$, $\epsilon_e=0.1$, and $p=2.3$. 
The burst is assumed to be located at a cosmological redshift $z=1$. 
The two different blast-wave evolutions shown in Figure~\ref{fig:dyn} 
yield significantly different light curves. 
The afterglow calculations are terminated when the RS arrives 
at the last shell ($\tau = 50$ s) in the ejecta.
Thus, from $\tobs \sim \mbox{few} \times 10^3~\mbox{s}$, 
the light curves are produced by high latitude emissions. 
} 
\label{fig:Rband}
\end{center}      
\end{figure}

\begin{figure}
\begin{center}
\includegraphics[width=11cm]{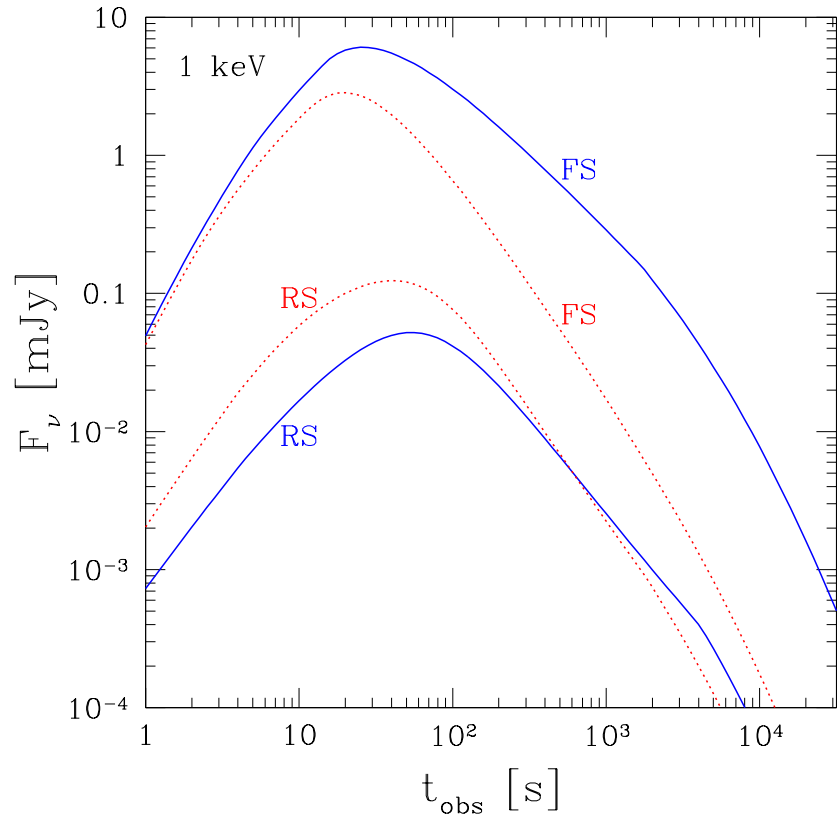}
\caption{
Same as in Figure~\ref{fig:Rband}, but in X-ray (1 keV) band.  
} 
\label{fig:Xray}
\end{center}      
\end{figure}

We also calculate the synchrotron emission from both the FS-shocked 
and RS-shocked regions. 
We make use of the standard prescription of microphysical 
parameters (e.g., Piran 2004): 
$\epsilon_e$ (fraction of the shock energy that goes to electron acceleration), 
$\epsilon_B$ (magnetic parameter), 
and $p$ (slope of the electron spectrum). 
We track the synchrotron emissivity of all shells on the blast; 
i.e., a radiative and adiabatic cooling of the electron spectrum is tracked for each shell, 
and an adiabatic evolution of shocked gas is tracked to give 
an evolution of the magnetic field for each shell.
The velocity and spherical curvature of the shells are also taken into account. 
A more detailed description will be presented elsewhere.
The resulting afterglow light curves are shown in Figure~\ref{fig:Rband} 
($R$ band) and Figure~\ref{fig:Xray} (1 keV).
We find that the two different blast-wave evolutions shown in Figure~\ref{fig:dyn} 
yield significantly different sets of light curves for the same example burst.
In particular, the light curves obtained with the pressure balance 
decrease earlier than those with the mechanical model, 
due to an energy loss seen in Figure~\ref{fig:etot_custom}.

A spherically symmetric formulation presented here becomes 
less accurate when the sideways expansion becomes important 
at late stages of the blast-wave evolution; 
i.e., $\Gamma \lesssim \theta_{\rm jet}^{-1}$. 
In their relativistic hydrodynamical simulations, 
Meliani et al. (2007) showed that a 2D jet-like model decelerates earlier than 
its 1D isotropic counterpart when thermally induced expansions lead to 
significantly high lateral speeds. On the other hand, 
Zhang \& MacFadyen (2009) showed that the sideways expansion is a very slow process 
and previous analytic works (Rhoads 1999; Sari et al. 1999) greatly overestimated 
the rate of the sideways expansion.

%
%

\section{Conclusion}

As the blast wave propagates, the strength of the RS wave exhibits a transition 
from non-relativistic to mildly relativistic or relativistic regime (or vice versa).
Thus, an EOS with a constant adiabatic index is not adequate for the RS-shocked region. 
We address that a more realistic EOS with a variable adiabatic index 
needs to be used for the gas in the RS-shocked region.

Following Mathews (1971), we consider a relativistic monoenergetic gas
and find its EOS. We show that there is only 4.8 \% of maximal difference 
in the quantity $\kappa$ (pressure divided by internal energy density)
when compared to a relativistic ideal gas.
Then we show that jump conditions of relativistic hydrodynamical shocks 
simplify significantly for the monoenergetic gas (see Section~\ref{section:shock}).
The simple form of jump conditions presented here 
is exact for a monoenergetic gas and applies to shocks of arbitrary strength 
(relativistic, mildly relativistic, or non-relativistic).
We emphasize that its usage is not to be restricted to GRB blast waves; 
it can be applied to other areas of relativistic hydrodynamical shocks.

Then we present a semi-analytic formulation for relativistic blast waves with 
a long-lived RS. We describe in detail a complete set of tools for finding 
a dynamical evolution of the blast wave for a very general class of explosions. 
The ambient medium can have an arbitrary radial profile, 
and the explosion ejecta can also be arbitrary as long as $\Gej^\prime(\tau) \leq 0$. 
We provide two different methods of finding dynamics of the blast wave: 
(1) customary pressure balance and 
(2) the mechanical model (Beloborodov \& Uhm 2006). 
Using a simple example model, we show that 
the pressure balance across the blast wave does not satisfy 
the energy conservation for an adiabatic blast wave; the total energy is 
decreased by a factor of 5 in the case of the example model.

The mechanical model does not assume a pressure balance or proportionality 
across the blast wave (neither $\pf=\pr$ nor $\pf/\pr={\rm const.}$ is assumed).
Instead, it finds the dynamics of the blast wave from a set of 
coupled differential equations that express the conservations of 
energy-momentum tensor and mass flux applied on the blast between the FS and RS.
Using the same example model, we show that the energy conservation is 
satisfied for the mechanical model as expected.

We also show that the two methods yield very different 
dynamical evolutions of the blast wave and, as a result, 
very different afterglow light curves. 
We conclude that the customary prescription of pressure balance 
poorly describes the dynamics of the blast wave with a long-lived RS 
and resulting afterglow light curves are inaccurate in a significant manner.


%
%

\acknowledgments

The author is grateful to Andrei M. Beloborodov 
for numerous helpful discussions and generous contributions to this work.
The author is also grateful to Robert Mochkovitch, Fr\'ed\'eric Daigne, and 
the anonymous referee for helpful comments to improve the manuscript.
This research was supported by 
the WCU program (R32-2009-000-10130-0) of NRF/MEST of Korea, 
the grant ``Research in Paris 2010/2011'' of the City Hall of Paris, 
and the French Space Agency (CNES).

%
%


\end{document}